\documentclass[fleqn,usenatbib]{mnras}

\usepackage[T1]{fontenc}

\DeclareRobustCommand{\VAN}[3]{#2}
\let\VANthebibliography\thebibliography
\def\thebibliography{\DeclareRobustCommand{\VAN}[3]{##3}\VANthebibliography}

\usepackage{graphicx}	
\usepackage{amsmath}	
\usepackage{amssymb}	

\usepackage{todonotes}

\usepackage{newtxtext,newtxmath}

\title[Evolution \& Structure of Embedded Clusters]{Early Evolution and 3D Structure of Embedded Star Clusters}

\author[Cournoyer-Cloutier et al.]{
Claude Cournoyer-Cloutier$^{1}$\thanks{E-mail: cournoyc@mcmaster.ca},
Alison Sills$^{1}$,
William E. Harris$^{1}$,
Sabrina M. Appel$^{2}$,
Sean C. Lewis$^{3}$,
\newauthor{ Brooke Polak$^{4,5}$,
Aaron Tran$^{6}$,
Martijn J.~C. Wilhelm$^{7}$,
Mordecai-Mark Mac Low$^{5,6,3}$,}
\newauthor{ Stephen L.~W. McMillan$^{3}$, Simon Portegies Zwart$^{7}$}
\\
$^{1}$Department of Physics and Astronomy, McMaster University, 1280 Main St W. Hamilton, L8S 4L8, Canada\\
$^{2}$Department of Physics and Astronomy, Rutgers University, 136 Frelinghuysen Road, Piscataway, NJ 08854, USA\\
$^{3}$Department of Physics, Drexel University, 3141 Chestnut Street, Philadelphia, PA 19104, USA\\
$^{4}$Zentrum f\"{u}r Astronomie, Institut f\"{u}r Theoretische Astrophysik, Universit\"{a}t Heidelberg, Albert-Ueberle-Stra{\ss}e 2, D-69120 Heidelberg, Germany\\
$^{5}$Department of Astrophysics, American Museum of Natural History, 200 Central Park W.
New York, NY 10024, USA\\
$^{6}$Department of Astronomy, Columbia University, 538 West 120th Street, New York, NY 10027, USA\\
$^{7}$Leiden Observatory, Leiden University, P.O. Box 9513, 2300 RA Leiden, The Netherlands
}

\date{Accepted 2023 February 16. Received 2023 February 03; in original form 2022 December 16}

\pubyear{2023}

\begin{document}
\label{firstpage}
\pagerange{\pageref{firstpage}--\pageref{lastpage}}
\maketitle

\begin{abstract}
We perform simulations of star cluster formation to investigate the morphological evolution of embedded star clusters in the earliest stages of their evolution.
We conduct our simulations with \textsc{Torch}, which uses the \textsc{Amuse} framework to couple state-of-the-art stellar dynamics to star formation, radiation, stellar winds, and hydrodynamics in \textsc{Flash}.  
We simulate a suite of 10$^4$~M$_{\odot}$ clouds at 0.0683 pc resolution for $\sim 2$~Myr after the onset of star formation, with virial parameters $\alpha=0.8, 2.0, 4.0$ and different random samplings of the stellar initial mass function and prescriptions for primordial binaries. 
Our simulations result in a population of embedded clusters with realistic morphologies (sizes, densities, and ellipticities) that reproduce the known trend of clouds with higher initial $\alpha$ having lower star formation efficiencies. 
Our key results are as follows: (1)~Cluster mass growth is not monotonic, and clusters can lose up to half of their mass while they are embedded. (2)~Cluster morphology is not correlated with cluster mass and changes over $\sim 0.01$~Myr timescales. (3)~The morphology of an embedded cluster is not indicative of its long-term evolution but only of its recent history: radius and ellipticity increase sharply when a cluster accretes stars. (4)~The dynamical evolution of very young embedded clusters with masses $\lesssim 1000$ M$_{\odot}$ is dominated by the overall gravitational potential of the star-forming region rather than by internal dynamical processes such as two- or few-body relaxation. 
\end{abstract}




\section{Introduction}
Most stars form within embedded clusters~\citep{Lada2003, PortegiesZwart2010}. They remain shrouded in their natal gas for a few megayears after the onset of star formation~\citep[see e.g.][for recent observations]{Kim2022}, while the cloud is still actively star-forming. Although most stars do not remain in bound star clusters for their whole lives, their formation and early evolution is shaped by the dense stellar environment in which they are born, which is in turn shaped by the interplay between gravity, turbulence, and stellar feedback. On smaller scales, stars also do not form in isolation: most stars form in multiple stellar systems~\citep[][and references therein]{Offner2022}, most often in binaries. Binaries are known to be dynamically important for cluster long-term evolution~\citep{Heggie1975, Hills1975}. Recent simulations by \citet{Torniamenti2021} further suggest that the presence of binaries impacts a cluster's structure over timescales of a few megayears after it has become free of gas. Despite their ubiquity, binaries in embedded clusters are seldom modelled numerically due the range of physical processes involved and the high numerical cost of modelling concurrently stellar dynamics on the scale of binaries and feedback processes impacting the gas in the embedded cluster. 

Simulations of star cluster formation show that star clusters assemble through the merging of smaller embedded clusters over a few megayears~\citep[e.g.][]{Fujii2012, Vazquez2017, Grudic2018, Howard2018, Chen2021}. \citet{Karam2022} have further shown that those mergers have an important impact on the boundedness of the stars and gas in the resultant cluster: some head-on collisions between clusters do not result in a single bound cluster, while there is mass loss and an increase in radius even in the successful mergers. The simulations conducted by~\citet{Karam2022} however do not account for the formation of new stars during cluster assembly. Recent work by~\citet{Dobbs2022}, which relies on star particles representing low-mass stellar populations or massive stars to model clusters, also reveals a more complex picture: clusters can not only merge, but also split. They also trace the mass and size of their clusters throughout their simulations, and find no clear correlation between mass and size. They however assume a spherical shape when measuring the size of their clusters, which is not the case for observed embedded clusters~\citep[e.g.][]{Kuhn2014}. Furthermore, neither of these recent suites of simulations include binaries, which are expected to influence stellar dynamics on the cluster scale, at least once the cluster becomes free of gas.

The virial parameter $\alpha$ of the star-forming cloud of gas, which describes the balance between the effects of self-gravity and turbulent support of the gas, is also important for cluster formation and evolution. For a spherical cloud, the virial parameter is defined as
\begin{equation}\label{eq:virial}
    \alpha = \frac{2T}{|U|} = \frac{5 \sigma^2 R}{GM}
\end{equation}
where $T$ is the kinetic energy of the cloud, $U$ is its gravitational potential energy, $\sigma$ is its velocity dispersion, $R$ is its radius, $M$ is its mass, and $G$ is the gravitational constant~\citep{Bertoldi1992}. Thus clouds with smaller $\alpha$ are more strongly bound, and $\alpha=1$ corresponds to virial equilibrium. Observed clouds in galaxies cover a large range of virial parameters, from $\alpha \lesssim 0.1$ to  $\alpha \gtrsim 100$~\citep{Kauffmann2013}. A cloud's virial parameter systematically affects its star formation efficiency (SFE) and cluster formation efficiency~\citep[CFE, ][]{Kruijssen2012, Howard2016}, with regions with higher $\alpha$ generally having lower SFE and CFE. 

In this work, we use numerical simulations to investigate the effects of stellar dynamics and cloud-scale hydrodynamics on the structure and evolution of embedded star clusters. To test the relative importance of stellar dynamics, we explore the impact of forming (or not forming) primordial binaries with different underlying populations, as binaries are known to play an important role in setting cluster structure in systems dominated by stellar dynamics~\citep[e.g.][]{Heggie1975, Fujii2011, Torniamenti2021}. To test the relative importance of cloud-scale hydrodynamics, we vary the cloud's initial virial parameter $\alpha$, which is known to have a strong effect on the CFE~\citep[e.g.][]{Howard2016}. 
We want to determine (1) whether cloud-scale hydrodynamics or stellar dynamics have the strongest impact on cluster structure (mass, size, and shape) and cluster formation efficiency and (2) how cluster structure evolves during the earliest stages of formation. 

In Section~\ref{sec:Methods}, we describe our numerical framework and our simulations. In Section~\ref{sec:Results}, we follow the evolution of the bulk properties of the stars in the simulation domain, we investigate the instantaneous properties of the clusters as a population, and we examine the assembly history of individual clusters; Section~\ref{subsec:results_time} contains the key results of the paper. In Section~\ref{sec:Discussion}, we discuss the broader implications of our findings. We summarize our results in Section~\ref{sec:Conclusion}.

\section{Methods}\label{sec:Methods}
\subsection{Numerical Framework}\label{subsec:Torch}
We use \textsc{Torch}~\citep{Wall2019, Wall2020, Cournoyer-Cloutier2021}, which relies on the \textsc{Amuse} framework~\citep{PortegiesZwart2009, Pelupessy2013, PortegiesZwart2013, amuse} to couple hydrodynamics to stellar dynamics, star and binary formation via sink particles, stellar evolution, and stellar feedback in the form of winds and radiation. \textsc{Torch} is optimized to investigate the effects of stellar and binary dynamics in young, gas-rich clusters, in particular stable multiple systems and dynamical short-range encounters between stars. We model the self-gravitating gas with the adaptive mesh refinement code \textsc{Flash}~\citep{Fryxell2000}. We use simultaneously two types of refinement criteria for our adaptive grid. We first require that the Jeans length be resolved by at least four resolution elements in order to avoid numerical fragmentation~\citep{Truelove1997, Federrath2010}. To improve stability, we also refine where the magnitude of the second derivative of the presssure, the temperature, the total energy or the internal energy is of the order of the sum of its gradients~\citep{Lohner1987, MacNeice2000}. Although \textsc{Flash} can evolve magnetic fields, we do not include them in our simulations due to their high computational cost. We treat gas dynamics with a Harten-Lax-van Leer Riemann solver~\citep{Miyoshi2005} and an unsplit (magneto)-hydrodynamics solver~\citep{Lee2013} with third-order piecewise parabolic method reconstruction~\citep{Colella1984}. We handle the gas self-gravity with a multigrid solver~\citep{Ricker2008} while we handle the gravitational attraction of the gas on the stars and vice-versa with a leapfrog scheme~\citep[][based on~\citeauthor{Fujii2007}~\citeyear{Fujii2007}]{Wall2019}.

On the stellar dynamics side, we handle long-range stellar dynamics with the direct N-body code \textsc{Ph4}~\citep{McMillan2012}, which uses a fourth-order Hermite predictor-corrector scheme~\citep{Makino1992}. For stable binary (and higher order) systems, resonant encounters and scattering, we use the \textsc{Amuse} module \textsc{Multiples}~\citep{amuse}, which itself uses the codes \textsc{smallN}~\citep{Hut1995, McMillan1996} and \textsc{kepler}~\citep[originally developed as part of \textsc{Starlab, }][]{PortegiesZwart1999, Hut2010}. 

Star formation takes place within sink particles that are treated as star factories. The details of the sink implementation are presented in~\citet{Wall2019} for single star formation and~\citet{Cournoyer-Cloutier2021} for binary formation. Briefly, a sink particle is formed when the local gas density and convergence criteria outlined in~\citet{Federrath2010} are satisfied. Once formed, it samples an initial mass function~\citep{Kroupa2001} between 0.08~M$_{\odot}$ and 150~M$_{\odot}$ to generate a list of stars to be formed, using a Poisson sampling method first tested by~\citet{Sormani2017} and implemented in \textsc{Torch} by~\citet{Wall2019}. Each star in the list is formed when the sink has accreted sufficient mass, in order to ensure quasi-local mass conservation. The sink must also sit in cold (< 100 K) gas to form stars. Stars are formed with a gas-to-star conversion efficiency of 100\%. The additive properties of the Poisson distribution ensure that the sampling for the full simulation domain reproduces the IMF, despite possible stochastic variations within individual clusters. The decoupling allows the stars to be handled by the N-body solver \textsc{Ph4}, which is fourth-order accurate, instead of the second-order leapfrog scheme used for sink particles. Although the formation of individual stars is unresolved in our simulations, stellar dynamics are followed self-consistently after star formation.

Stars with masses above 7 M$_{\odot}$ inject radiative and momentum feedback on the grid. The details of the feedback implementation are presented in~\citet{Wall2020}. The far ultraviolet (between 5.6~eV and 13.6~eV) and ionizing (above 13.6 eV) radiative feedback is implemented within \textsc{Flash} as a modified version of the adaptive ray-tracing module \textsc{Fervent}~\citep{Baczynski2015}. The total and average photon energy are calculated for each star from the surface temperature and mass obtained from stellar evolution, which is performed with \textsc{SeBa}~\citep{PortegiesZwart1996}. All radiative feedback heats the gas. Massive stars further provide feedback in the form of momentum-driven winds with mass loss rates based on~\citet{Vink2000}. Radiative cooling of the gas from atomic and molecular lines and dust is included~\citep{Wall2019}. 

\subsection{Simulations and Star Formation Prescriptions}\label{subsec:Simulations}

\begin{table}
    \centering
    \begin{tabular}{lclc}
        Name \, \, & \, \, $\alpha$ \, \, & \, \, Primordial binaries \, \, &  Random seed  \\
        \hline
        B-P0 & 0.8 & Field distribution & Default \\
        B-P1 & 0.8 & 10\% random pairing & Seed 1 \\
        B-P2 & 0.8 & 100\% random pairing & Seed 2 \\
        B-P3 & 0.8 & Field distribution for $M < 0.6 M_{\odot}$ & Seed 3 \\
         &  & and no close massive binaries &  \\
        \hline
        S-R0 & 0.8 & None & Default \\
        S-R1 & 0.8 & None & Seed 4 \\
        S-R2 & 0.8 & None & Seed 5 \\
        S-R3 & 0.8 & None & Seed 6 \\
        \hline
        B-V2 & 2.0 & Field distribution & Default \\
        S-V2 & 2.0 & None & Default \\
        B-V4 & 4.0 & Field distribution & Default \\
        S-V4 & 4.0 & None & Default \\
        \hline
    \end{tabular}
    \caption{Overview of simulations' initial conditions and star formation prescriptions. $\alpha=2T/|U|$ denotes the virial parameter; bound clouds have $\alpha < 2$ and unbound clouds have $\alpha > 2$. The prescriptions for primordial binaries are outlined in Appendix~\ref{secA:binaries}.}
    \label{tab:sims}
\end{table}

We conduct a total of 12 simulations, summarized in Table~\ref{tab:sims}. All simulations are initialized from a spherical, turbulent cloud of neutral dense gas with a mass of $10^4$ M$_{\odot}$ and a radius of 7~pc in a cubic box of side 17.5~pc, following the model used in~\citet{Cournoyer-Cloutier2021}. The mean gas surface density is 50~M$_{\odot}$ pc$^{-2}$. Those values are consistent with a typical cloud in the Solar neighbourhood~\citep{Chen2020}. The cloud follows a Gaussian density profile with a central density 8.75~x~10$^{-22}$ g cm$^{-3}$ and temperature 20.64~K, and sits in a warm neutral medium with density 2.18~x~10$^{-24}$ g cm$^{-3}$ and temperature 6.11 x $10^{3}$~K. These values were chosen to ensure pressure and thermal equilibrium between the cloud and surrounding medium. The free-fall time for the cloud is 1.45~Myr. The gas follows an adiabatic equation of state with $\gamma = 5/3$, although radiative cooling maintains the dense neutral gas almost isothermal. We adopt the same gas spatial resolution of 0.0683~pc at the maximum refinement level and density threshold for the formation of sink particles of 3.82~x~10$^{-21}$ g cm$^{-3}$ as used in~\citet{Cournoyer-Cloutier2021}. 

We consider four different prescriptions for binaries (described in Appendix~\ref{secA:binaries}), in addition to models without primordial binaries. Our models with primordial binaries span a range of mass-dependant binary fractions, mass ratios, and orbital periods. We stress that the details of those prescriptions are not the focus of this paper -- rather, we test diverse models for primordial binaries to fully explore the impact that a change in stellar dynamics has on embedded cluster structure and evolution. Binaries can also form dynamically, and the properties of primordial and dynamically-formed binaries will be modified by dynamics over the course of our simulations. We refer the interested reader to~\citet{Cournoyer-Cloutier2021} for a detailed discussion of the effects of dynamical interactions on the initial population of binaries.

Eight of our 12 simulations (with names starting with B-P and S-R) are initialized with a virial parameter $\alpha = 2T/|U| =0.8$. The gas is initially gravitationally bound and its collapse is expected to result in abundant star formation. The gas initial conditions, including the random turbulent field, are identical for those 8 simulations. We also perform simulations with larger virial parameters, of $\alpha=2.0$ and $\alpha=4.0$. We perform pairs of simulations with our fiducial prescription for binaries and our single-stars only prescription (both with the default random seed) for both these models, and label them B-V2, S-V2, B-V4 and S-V4. Those initial conditions are set up by scaling up the gas velocities in each cell of the initial conditions for the $\alpha=0.8$ runs. We therefore increase $\alpha$ but conserve the direction of motion of the gas in each cell. 

Beyond the binary prescriptions and virial parameters, we also vary the random seed used to sample the initial mass function and to form binaries, which sets the masses of the stars and the order in which they form.  Our simulations labelled with the random seed \textit{default} all use the same random seed; the other simulations all use different random seeds. We use different random seeds to ensure our general conclusions are not affected by the stochastic formation of massive stars. We have shown in~\citet{Lewis2022} that early-forming massive stars can promote the formation of smaller, isolated clusters, and prevent the formation of massive clusters. By using different random samplings of the IMF, we can verify that our conclusions are not drawn from a single, extreme case, in which the formation times and masses of the massive stars providing radiative and mechanical feedback would be atypical.

\subsection{Cluster Identification}\label{subsec:DBSCAN}
\begin{figure*}
    \centering
    \begin{minipage}{.48\textwidth}
        \centering
        \includegraphics[width=\textwidth, clip=True, trim=8cm 4cm 4cm 6cm]{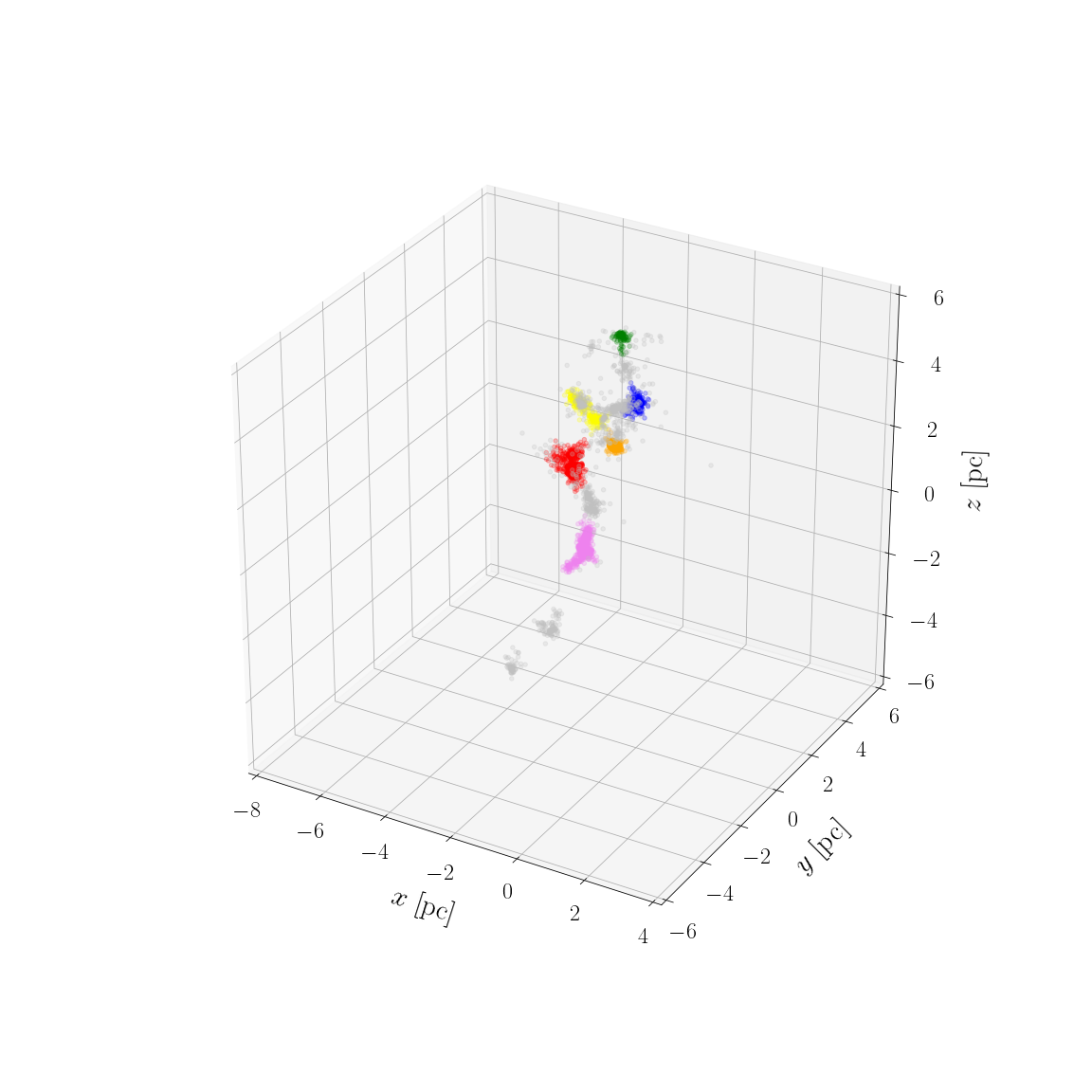}
    \end{minipage}
    \begin{minipage}{.48\textwidth}
        \centering
        \includegraphics[width=\textwidth, clip=True, trim=8cm 4cm 4cm 6cm]{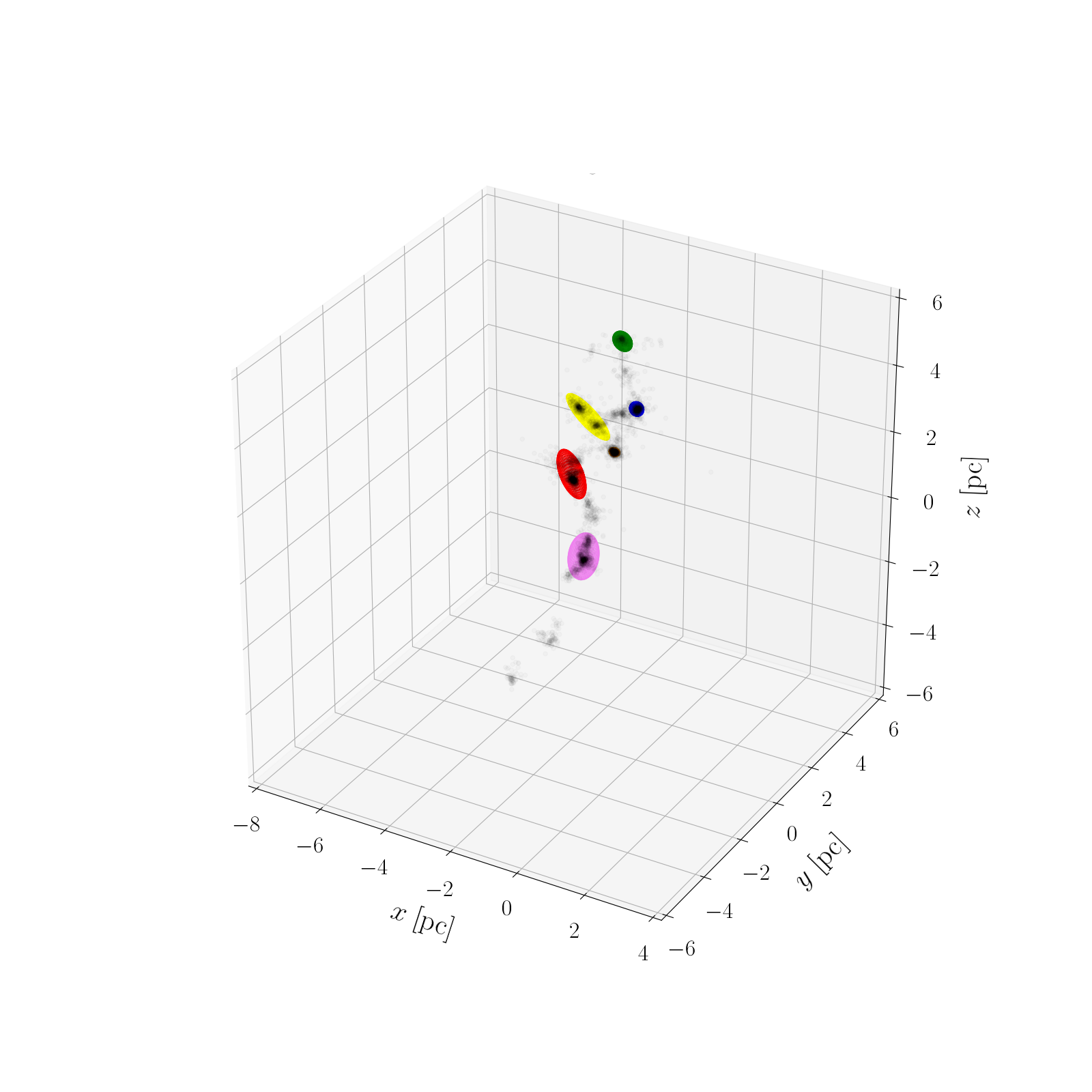}
    \end{minipage}
    \caption{\textbf{Left:} Example of 3D spatial clustering of the stars in S-R0 at the last snapshot, 2 Myr after the onset of star formation. Member stars for each cluster with at least 100 bound members are shown in a given colour -- blue, green, yellow, orange, red, or pink -- while unclustered stars are shown in grey. \textbf{Right:} Example of ellipsoids enclosing 90\% of the cluster mass for the bound clusters identified in the last snapshot of S-R0, 2 Myr after the onset of star formation. The colours of the ellipsoids match those of the members stars identified on the left.}
    \label{fig:DBSCAN}
\end{figure*}

Most of our simulations have reached 2 Myr after the onset of star formation, and snapshots are written every 0.01 Myr. We inspect all snapshots in our simulations for clusters, which we identify from a combination of spatial clustering and boundedness. We initially select clusters with DBSCAN~\citep{Ester1996, scikit-learn} based on the positions of the stars. We require each cluster star to have five neighbours~\citep[following][for three-dimensional data]{Sander1998}, which are other stars within a user-determined distance. For our analysis, we fix this distance to the sink accretion radius, 0.17~pc. Following our initial identification of the clusters, we perform a boundedness check on the stars with respect to their associated cluster. For each star, we calculate the gravitational potential energy from the local gas gravitational potential (including the sink particles) and the potential from the cluster's stars. We also calculate the stars' kinetic energy in the cluster's centre of mass frame. We remove stars with positive total energy (i.e. unbound stars) from the cluster. After this boundedness check, clusters that have at least 100 members are saved for subsequent analysis. An example of the clusters satisfying our clustering, boundedness and minimum membership criteria in a given snapshot is shown in the left panel of Figure~\ref{fig:DBSCAN}.

\subsection{Cluster Structure}\label{subsec:ellipsoids}
Once clusters are identified, it is useful to describe their size, which in turn requires us to measure their shape. Observational studies have used respectively ellipses~\citep[e.g.][]{Kuhn2014, Zhai2017} and ellipsoids~\citep[e.g.][]{Pang2021} to describe the 2D and 3D shapes of embedded or open clusters. We similarly use 3D ellipsoids to describe the shape of some inner fraction of the stellar distribution in an individual cluster -- here, we use 50\% and 90\% mass ellipsoids, as proxies for the 50\% and 90\% Lagrangian radii. We use the fact that any distribution of points can be described by an inertial ellipsoid that shares its rotational properties about its principal axes~\citep[see e.g.][]{GoldsteinCM}. This technique has been used previously in astrophysics to describe the 3D shape (or projected 2D shape) of dark matter halos in cosmological simulations~\citep[see e.g.][]{Velliscig2015, Thob2019, Hill2021, Reina-Campos2022}. We show the 90\% mass ellipsoids for the last snapshot of S-R0 (the same example as for the clustering plot) in the right panel of Figure~\ref{fig:DBSCAN}. We present the details of our fitting routine in Appendix~\ref{secA:inertia}. An example of the 50\% and 90\% mass ellipsoids for an individual cluster identified in our simulations is also provided in Figure~\ref{fig:sphere_ellipsoid_example}, and the spherical half-mass and 90\% Lagragian radii are provided for comparison. 

We use our fitted ellipsoids to define a proxy for the radius, to compare our clusters to established mass-radius relations. We do so by taking the geometric mean of the semi-major, intermediate, and semi-minor axes $a$, $b$ and $c$ to define a characteristic radius
\begin{equation}\label{eq:characteristic_radius}
    \tilde{r} = (abc)^{1/3}
\end{equation}
which is similar to what is done in observational studies~\citep[e.g.][]{Kuhn2014} to define sizes for elliptical 2D clusters.
We can define such a radius for any enclosed mass fraction, and therefore for any Lagrangian radius. 
To quantify how non-spherical a cluster is, we define an ellipticity~\citep[see e.g.][]{Kuhn2014} 
\begin{equation}\label{eq:ellipticity}
    \epsilon = \frac{a-c}{a}
\end{equation}
that depends on the ratio between the semi-minor and semi-major axes. A spherical cluster has an ellipticity $\epsilon = 0$ while a very elongated cluster has an ellipticity $\epsilon \rightarrow 1$. With Equations~\ref{eq:characteristic_radius} and \ref{eq:ellipticity}, we characterize the size and shape of individual clusters at each snapshot in our simulations. 

\subsection{Cluster History}\label{subsec:history}
We follow the evolution of individual clusters throughout the simulations. For each cluster identified in the last snapshot of a simulation, we trace back its main progenitor in earlier snapshots by identifying the cluster sharing the largest fraction of its stellar mass in the previous snapshot. We also look for clusters that are present at earlier times but are no longer present in the last snapshot. We allow for clusters to be missing in some checkpoints (for example, if a cluster with 100 bound members loses one star, then forms one or more later on) but require a cluster to be present over at least 0.1~Myr to trace its history. In practice, this means that a cluster can be used in our analysis of cluster populations (e.g. Sections~\ref{subsec:results_overview} and~\ref{subsec:results_clusters}) without being used in our analysis of cluster histories (Section~\ref{subsec:results_time}) if it survives for less than 0.1~Myr. 

We use our results on cluster histories in three main ways. First, we track the evolution of the mass, size, and shape of individual clusters to investigate the presence of evolutionary trends. Second, we investigate the relative contributions of cluster mergers, the accretion of unclustered stars, and new star formation to the build-up of our clusters during the first $\sim$ 2 Myr after the onset of star formation. Third, we evaluate what proportion of cluster stellar mass is lost over the same time. Those relative contributions are not final, as the clusters are still growing in mass at the end of the simulations. They however give us a picture of the variations in cluster history during the early stages of embedded cluster evolution.

We rely on the tags given to star particles to follow the assembly of individual clusters. Between two subsequent snapshots in which a cluster is identified, we identify all new star particles and all star particles that left the cluster. For new star particles, we verify whether they were present in the previous snapshot (either in another cluster or as unclustered stars). If they were not present in the previous snapshot, we consider them to be newly-formed stars, and treat them as having formed in the cluster. If they were present, we consider them as accreted stars. Star particles that have left the cluster are recorded as lost stars. For accreted and lost stars, we ensure that there is no double-counting, which could occur for example if a merger is unsuccessful or if a cluster splits. To evaluate cluster assembly, retained stars are therefore treated as formed in the cluster, accreted, or lost, if they respectively fulfill the following criteria:
\begin{enumerate}
    \item Stars are considered \textit{formed in the cluster} if they were not present (as a clustered or unclustered star) in the snapshot before they are identified as a cluster member, and are present in the cluster in the last snapshot in the simulation. Some of the stars complying with these criteria may have been lost and then re-accreted.
    \item Stars are considered \textit{accreted} if they were present in another cluster or as an unclustered star in the snapshot before they are identified as a cluster member, and are present in the cluster in the last snapshot in the simulation. Such stars may also have been lost and then re-accreted.
    \item Stars are considered \textit{lost} if they were present in the cluster at any earlier snapshot, and are not present in the cluster in the last snapshot in the simulation. Such stars may also have been lost, re-accreted, and then lost again.
\end{enumerate}
The stars that were cluster members when the cluster was first identified are treated separately to avoid artificially driving up the formed or accreted fractions in low mass clusters. We record the composition of the cluster at the end of our simulations (i.e. the mass in initially-present, formed, and accreted stars), as well as the mass lost throughout the history of the cluster.

\section{Results}\label{sec:Results}
We structure our results in three subsections, corresponding to three different approaches to analyzing our simulations. In Section~\ref{subsec:results_overview}, we summarize the evolution of the full simulation by tracking properties such as the star formation rate (SFR) and the clustered stellar mass. In Section~\ref{subsec:results_clusters}, we track the mass, size, and morphology of the identified clusters as a population, and compare them to observations of Galactic clusters. In Section~\ref{subsec:results_time}, we explicitly follow the evolution of the clusters throughout the simulations by tracking how they assemble their mass and how their morphologies change. 

\subsection{Overview: Properties of the Full Simulation Domain}\label{subsec:results_overview}

\begin{figure}
    \centering
    \includegraphics[width=\linewidth, clip=true, trim=1.5cm 1cm 1cm 1cm]{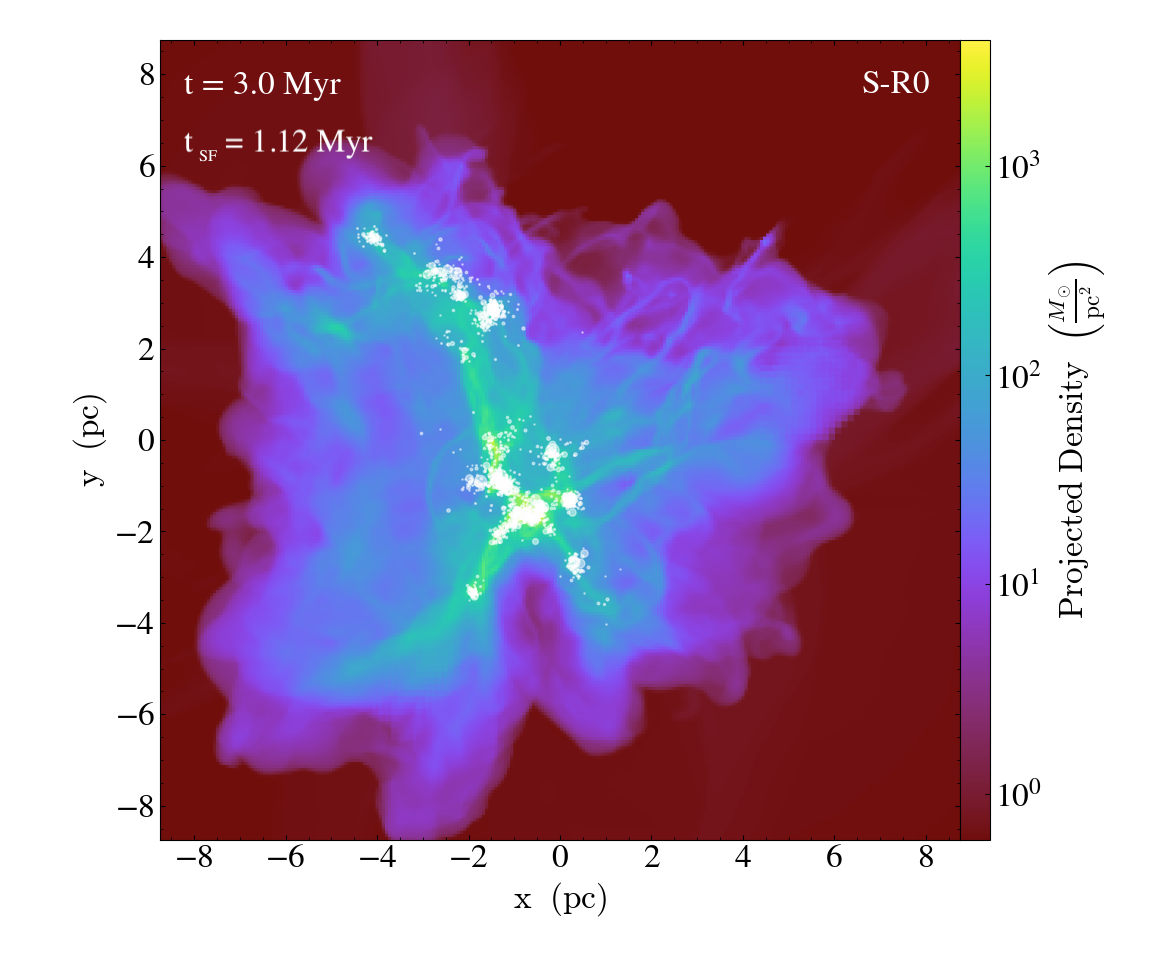}
    \includegraphics[width=\linewidth, clip=true, trim=1.5cm 1cm 1cm 1cm]{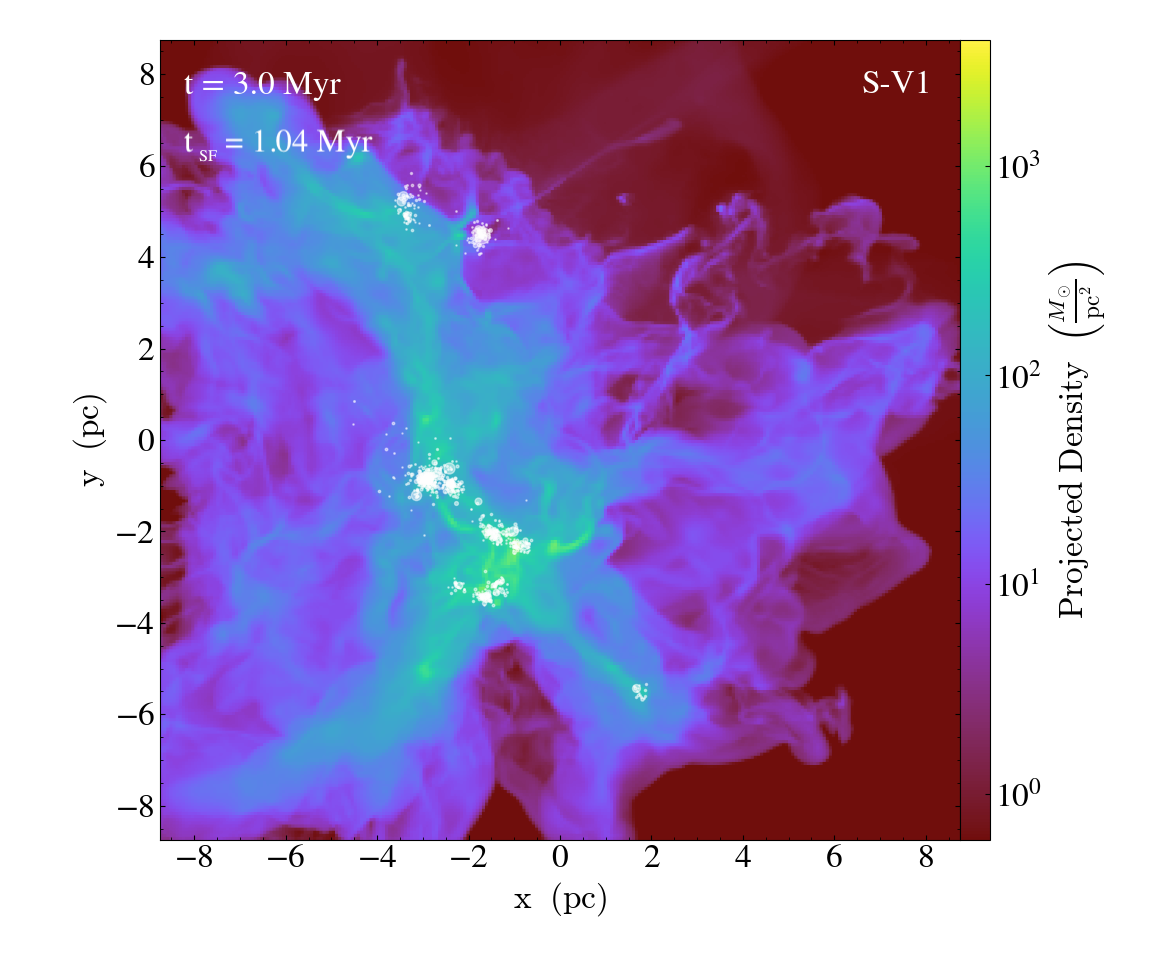}
    \includegraphics[width=\linewidth, clip=true, trim=1.5cm 1cm 1cm 1cm]{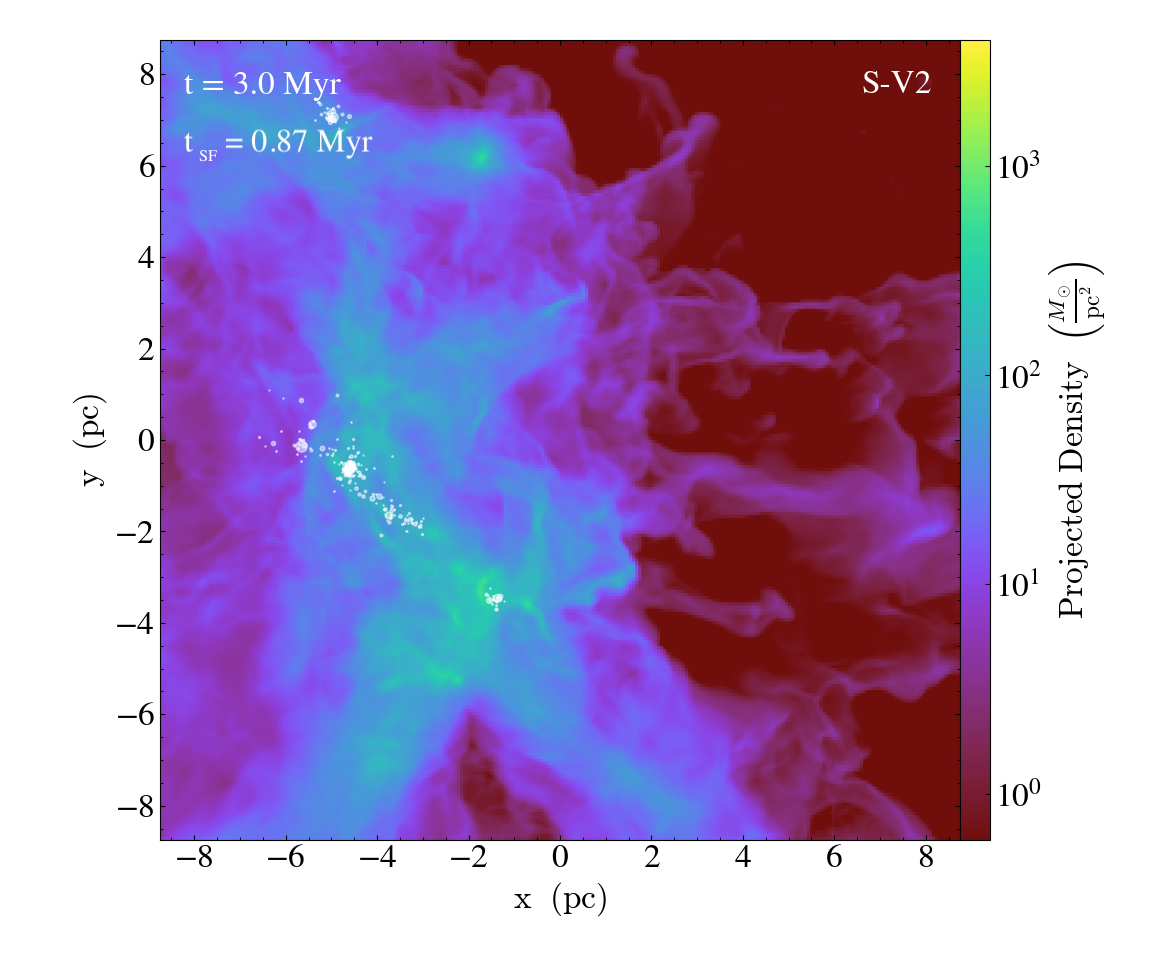}
    \caption{Gas surface density along the $z$ axis for simulations initialized with the different virial parameters $\alpha$ (from top to bottom, $\alpha$=0.8, 2.0, 4.0), 3.0 Myr after the start of the simulation. Star formation begins at a time $t_{SF}$ (labelled for each frame) after the start of the simulation. All three simulations form single stars only and use the same random seed for star formation. Stars are shown in white, with a marker size proportional to the star's mass.}
    \label{fig:gas}
\end{figure}

We first look at the global properties of the simulations. The starkest differences are between simulations with different initial virial parameters $\alpha$. This is already obvious from the plots of the gas column density presented in Figure~\ref{fig:gas}. The three simulations shown in the figure have the same star formation model (single stars only, default random seed) but are initialized with virial parameters of respectively $\alpha=0.8$, 2.0 and 4.0. Some features in the gas (such as the inverted Y shape made by the densest gas) persist across the three plots, but the gas morphology is nonetheless obviously different in the three simulations. In particular, the gas is less centrally concentrated and closer to the edges of the domain in the simulations with larger virial parameters.

\begin{figure}
    \centering
    \includegraphics[width=\linewidth, clip=true, trim=0 5cm 2cm 5cm]{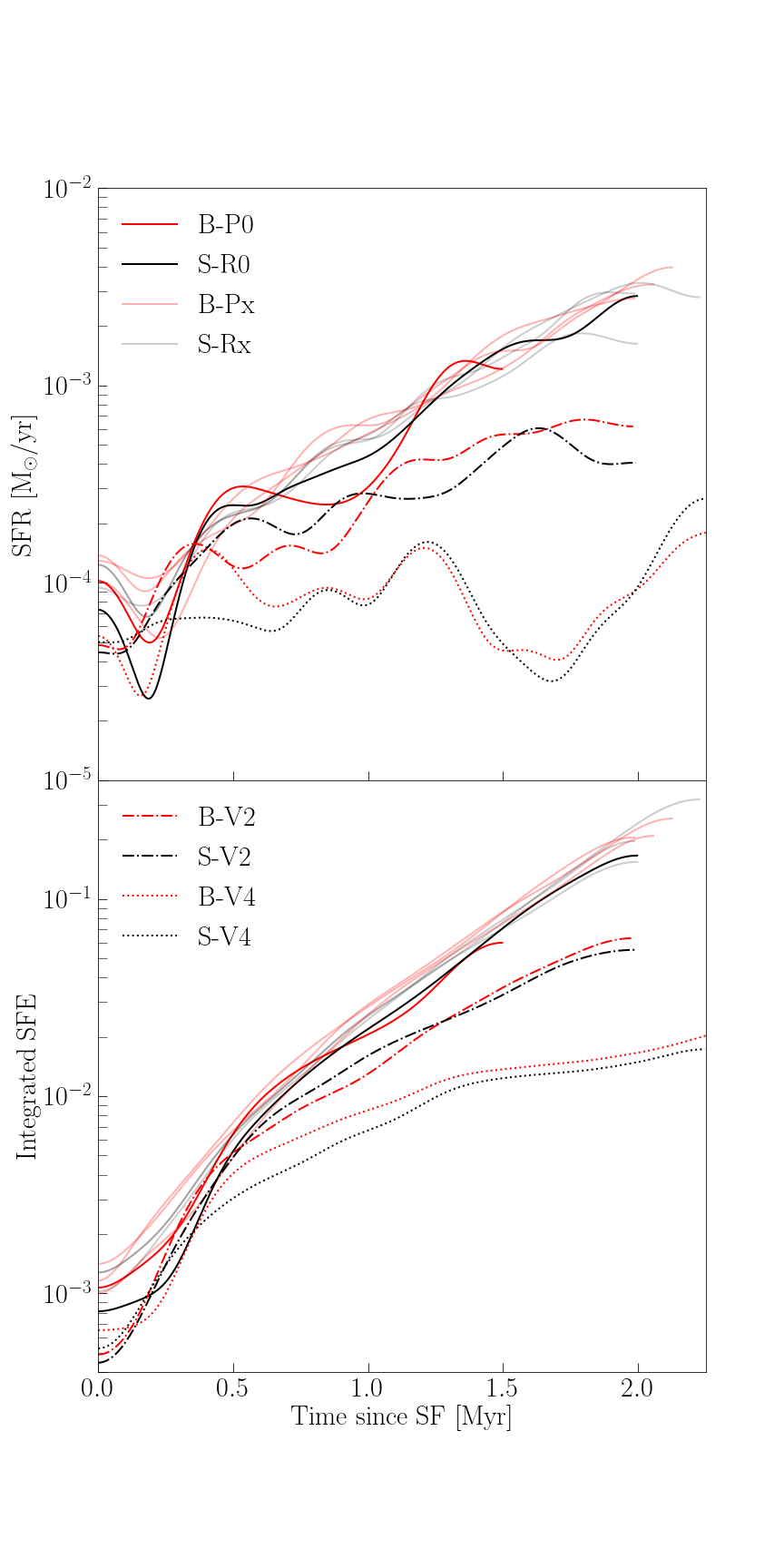}
    \caption{SFR (top) and integrated SFE (bottom) plotted against the time since the onset of star formation for the different simulations, smoothed over 0.1 Myr using a Gaussian filter. Simulations with primordial binaries are shown in red and simulations with single stars only are shown in black. Transparent red and grey are used for the runs that do not use the default random seed (respectively B-P1, B-P2, and B-P3, and S-R1, S-R2, and S-R3). Solid lines are used for simulations with $\alpha=0.8$, dashed-dotted lines for simulations with $\alpha=2.0$, and dotted lines for simulations with $\alpha=4.0$. Simulations with different $\alpha$'s display different general trends but simulations with the same $\alpha$ and different stellar populations do not.}
    \label{fig:sfr}
\end{figure}

Those morphological differences naturally give rise to differences in the SFR. The SFR and integrated SFE (mass of all formed stars divided by the initial gas mass) for the different simulations are plotted against time since the onset of star formation (SF) in Figure~\ref{fig:sfr}. We use a Gaussian filter with a kernel width of 0.1~Myr to smooth both the SFR and the SFE, in order to remove instantaneous peaks in the SFR caused by the formation of individual massive stars. By the time we stop the simulations, the SFR and the SFE are both about half an order of magnitude larger in our simulations with the fiducial $\alpha=0.8$ than in the simulations with $\alpha=2.0$, and more than an order of magnitude larger than in the simulations with $\alpha=4.0$. The different prescriptions for binary formation and the choice of random seed for star formation do not systematically affect the SFR or the SFE. They however cause scatter, which is smaller than the systematic effects associated with variations in $\alpha$. 

Simulations with higher virial parameters begin forming stars earlier than simulations with lower virial parameters. The first stars form respectively at $t_{SF} = 1.12, 1.04, 0.87$ Myr in simulations with $\alpha =0.8, 2.0, 4.0$. This is consistent with our expectations: turbulence both promotes star formation -- in leading to an earlier onset of SF -- and prevents it -- in lowering the SFR~\citep{Ballesteros-Paredes2007} but its net effect is to decrease the SFR~\citep{MacLow2004}. We also note that the clusters (and the stars) in our simulations with $\alpha=0.8$ tend to form along a linear chain of width $\sim$ 1 pc (see Figure~\ref{fig:DBSCAN}). This is similar to the complexes of embedded clusters in DR 21, NGC 2264, NGC 1893, NGC 6334, and the Carina Nebula observed by~\citet{Kuhn2014} in the MYStIX survey.

\subsection{Average Properties of Individual Clusters}\label{subsec:results_clusters}

We now turn our attention to the properties of embedded clusters identified in our simulations as a population. For this section, we use all clusters identified at all times in our simulations and measure their masses, sizes, and ellipticities. 

\begin{figure}
    \centering
    \includegraphics[width=\linewidth, clip=True, trim =0 1.5cm 2cm 2cm]{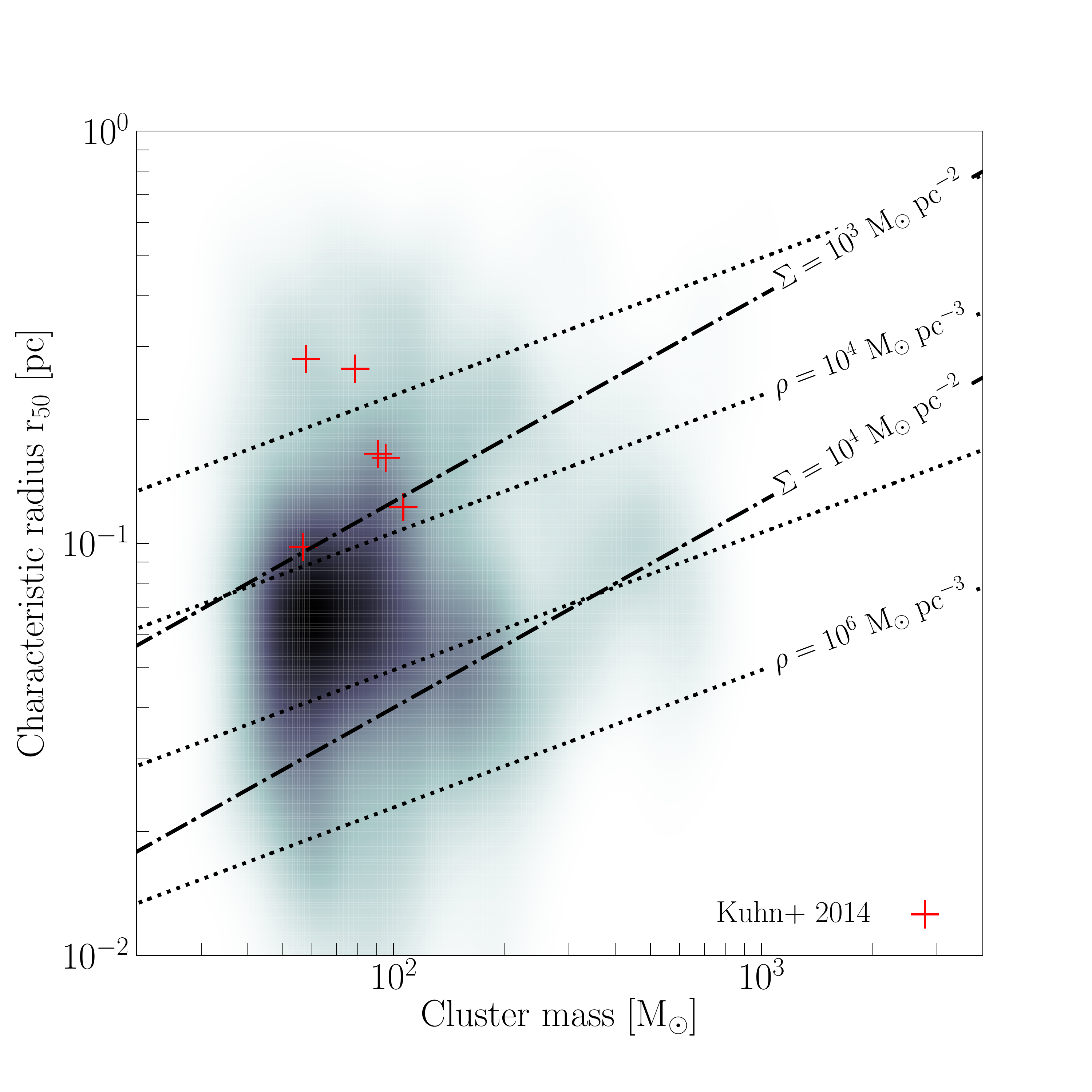}
    \caption{Distribution of characteristic radius $r_{50}$ against cluster mass, for all clusters identified in each snapshot of our simulations. Brightness decreases linearly with increasing density in parameter space. The dotted lines denote constant densities and the dashed-dotted lines denote constant surface densities. A few high density clusters with masses $\sim$ 100 M$_{\odot}$ and radii <~0.01~pc (discussed in-text) lie beyond the limits of the plot. Six deeply embedded clusters with at least 100 members from the MYStIX survey~\citep[][]{Kuhn2014} are shown in red for comparison.}
    \label{fig:mass-size}
\end{figure}

In Figure~\ref{fig:mass-size}, we present a mass-radius plot for all individual clusters in our simulations, where the characteristic radius $r_{50}$ for the 50\% ellipsoid is calculated with Equation~\ref{eq:characteristic_radius} and the cluster mass is obtained from the sum of the masses of all stars identified as cluster members. The diagonal lines denote lines of constant mass density or surface density. We also show in Figure~\ref{fig:mass-size} the characteristic radii of six deeply embedded clusters  with median X-ray energy in the 0.5-8.0~keV band above 2.0~keV and at least 100 members from the MYStIX survey~\citep{Kuhn2014}. Median X-ray energy is a proxy for extinction, and therefore anti-correlated with cluster age; the six selected clusters are expected to be the best match to our simulated cluster population in age and mass. We calculate the characteristic radii of the observed clusters from 
\begin{equation}\label{eq:projected_radius}
    \tilde{r} = (ab)^{1/2}
\end{equation}
where $a$ and $b$ are respectively the semi-major and semi-minor axes of the projected ellipses. We estimate the mass $M$ of the observed clusters from their star counts, using 
\begin{equation}
    M = 0.485 M_{\odot} \, N
\end{equation}
where $N$ denotes the star count, and the slope is obtained by fitting the mass against the star count for our simulated clusters. Most identified embedded clusters have masses around 100 $M_{\odot}$, characteristic half-mass radii around 0.05 pc, and therefore densities (calculated within the characteristic half-mass radius) between 10$^4$ and 10$^5$ M$_{\odot}$ pc$^{-3}$. The most massive clusters have densities around 10$^5$ M$_{\odot}$ pc$^{-3}$. This is approximately the same density as the Arches cluster~\citep{Serabyn1998}, which has a similar age of $\sim$ 2 Myr but a mass of a few $10^4$ M$_{\odot}$, about two orders of magnitude larger than our clusters. We therefore conclude that our clusters have densities comparable to the upper limit of observed densities in young Galactic clusters. 

Using the two-sample Kolmogorov-Smirnov (KS) test, we find no statistically significant difference in the masses, radii, or densities of the clusters identified in simulations with and without primordial binaries. This supports our earlier conclusion that there are no structural differences in clusters with and without primordial binaries over the timescales spanned by our simulations. Similarly, the clusters formed in simulations with different virial parameters cover similar regions in mass-radius space. 

We also find clusters with unphysically high densities (above 10$^{10}$ M$_{\odot}$ pc$^{-3}$) within $r_{50}$, which is often due to a single star accounting for $\gtrsim 30$ \% of the cluster's mass. One star may contribute to up to $\sim 50$\% of the cluster's mass: an extreme example is a $\sim$ 90 M$_{\odot}$ star in a $\sim$ 200 M$_{\odot}$ cluster, in S-R2. This skews the mass density to much higher than that of observed clusters but the number density remains reasonable. The clusters discussed here are still actively forming, and we expect them to grow via the formation of new stars and mergers with other clusters before star formation halts; the massive star discussed above is therefore expected to become part of a more massive cluster or to be lost as a runaway star.

\begin{figure}
    \centering
    \includegraphics[width=\linewidth, clip=True, trim=0cm 3cm 1cm 4.5cm]{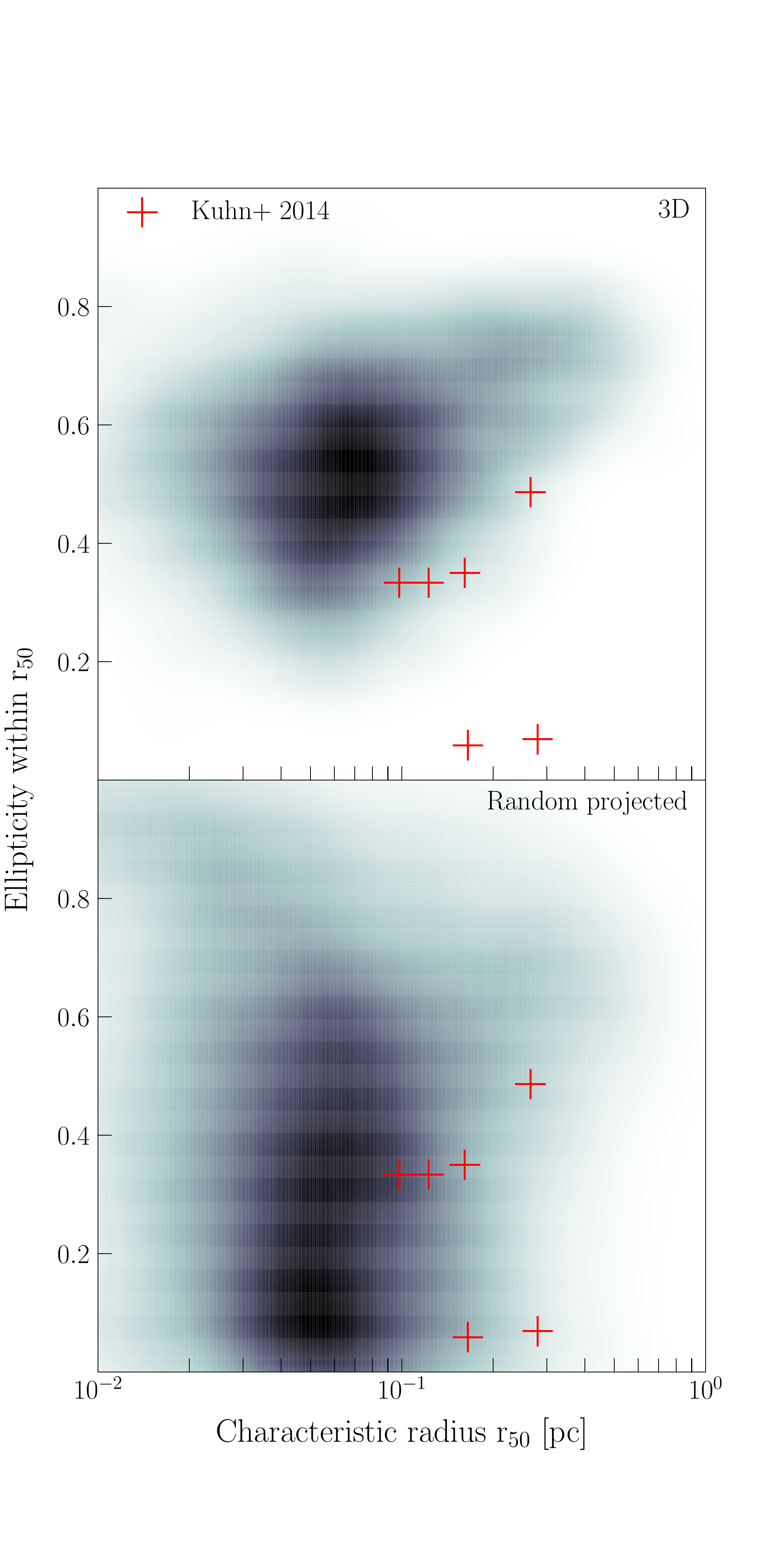}
    \caption{Ellipticity (top) and ellipticity of the projected ellipses along a random direction (bottom), against effective cluster radius $r_{50}$. Density in parameter space increases linearly with decreasing brightness. Six deeply embedded clusters with at least 100 members from the MYStIX survey~\citep[][]{Kuhn2014} are shown in red for comparison.}
    \label{fig:radius_ellipticity}
\end{figure}

We plot cluster ellipticity against characteristic radius $r_{50}$ in Figure~\ref{fig:radius_ellipticity}. In the top panel, the ellipticity shown is that of the ellipsoid enclosing 50\% of a cluster's mass, calculated with Equation~\ref{eq:ellipticity}. We compare the radius-ellipticity distribution to that of the same six deeply embedded clusters with at least 100 members from the MYStIX survey~\citep{Kuhn2014}. For the observed embedded clusters, we calculate the characteristic radius from Equation~\ref{eq:projected_radius}. Given the apparent mismatch between the simulated and observed clusters, we explore projection effects. To complete this more robust comparison to observations, we calculate the size and ellipticity from 2D projections of the simulated clusters' shapes and present them in the bottom panel. Each 3D ellipsoid is projected along a randomly-selected axis, and the semi-major and semi-minor axes $\tilde{a}$ and $\tilde{b}$ of the projected ellipse are used to calculate the characteristic radius and ellipticity respectively from Equations~\ref{eq:projected_radius} and~\ref{eq:ellipticity}. When accounting for projection effects, we find that our simulated clusters have ellipticities similar to those of the deeply embedded objects in the MYStIX sample, although our simulated clusters tend to have smaller radii. \citet{Kuhn2014} however find that the sizes of embedded clusters in their sample are positively correlated with cluster age. For their sub-sample of very deeply embedded objects -- which are the most comparable in age to our simulated clusters but not limited in star count -- they find an average projected radius of 0.04 pc, which is in good agreement with our simulated clusters. 

Our simulated embedded clusters have realistic masses, sizes, densities, and ellipticities: conclusions drawn from the study of their evolution can therefore inform our understanding of observed embedded clusters. We further note that there are no systematic differences in the structural properties of our simulated embedded clusters as a population regardless of the presence of primordial binaries or the choice of initial virial parameter for the star-forming cloud. 

\subsection{Time Evolution of Individual Clusters}\label{subsec:results_time}
We now investigate the evolution of individual clusters throughout the simulations. We find no individual cluster satisfying our membership and boundedness criteria that survived more than 0.1~Myr and then merged with another cluster. We however find that some clusters acquire more than $\sim 100$ M$_{\odot}$ due to accretion. Two processes contribute to this accretion budget without being registered as mergers. First, clusters satisfying our minimum membership and boundedness criteria may be accreted less than 0.1~Myr after they are first detected, and so before their histories are tracked. Second, groups of stars with fewer than 100 bound members -- that are not recorded as clusters -- may be accreted. We further find six examples of clusters splitting from an already-formed cluster, four of which survived for more than 0.1 Myr (the other two are identified because they are present in the last snapshot of the simulation). 

\begin{figure}
    \centering
    \includegraphics[width=\linewidth, clip=True, trim=20cm 0cm 120cm 0cm]{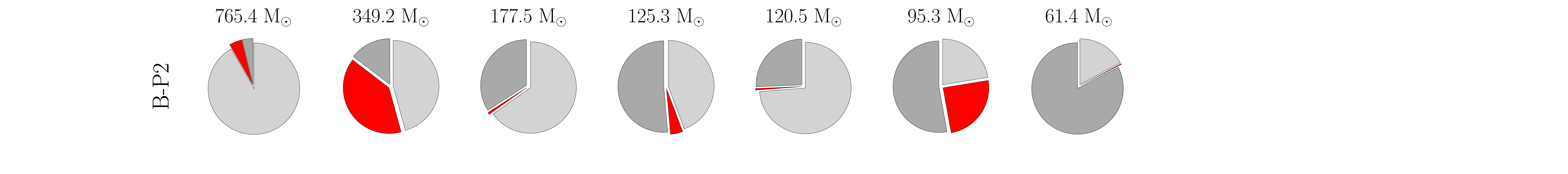}
    \includegraphics[width=\linewidth, clip=True, trim=20cm 0cm 120cm 0cm]{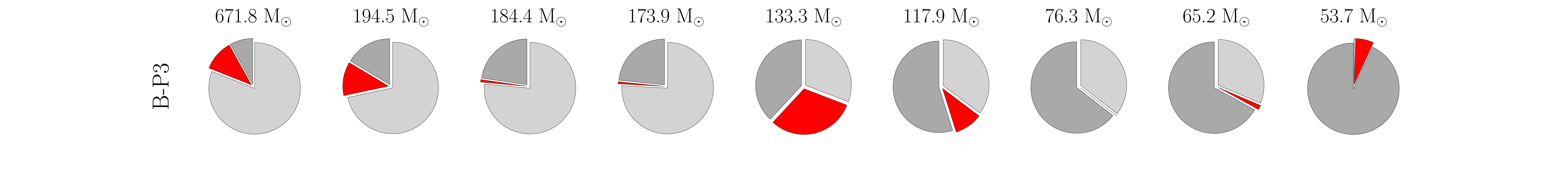}
    \includegraphics[width=\linewidth, clip=True, trim=20cm 0cm 120cm 0cm]{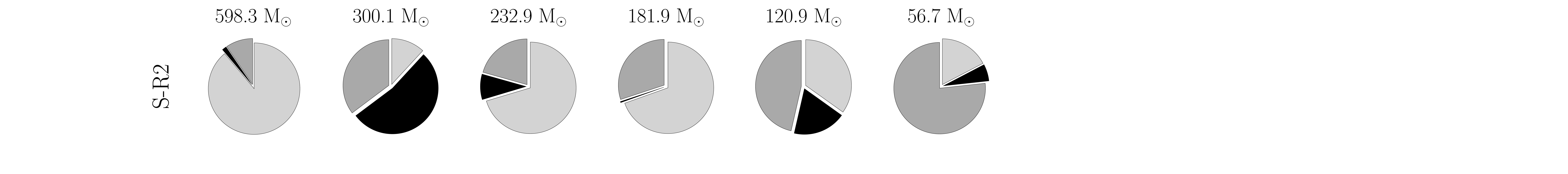}
    \includegraphics[width=\linewidth, clip=True, trim=20cm 0cm 120cm 0cm]{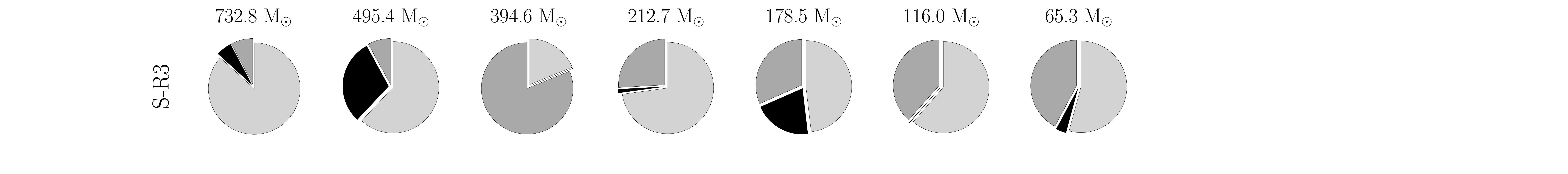}
    \includegraphics[width=\linewidth, clip=True, trim=0cm 0cm 0cm 0cm]{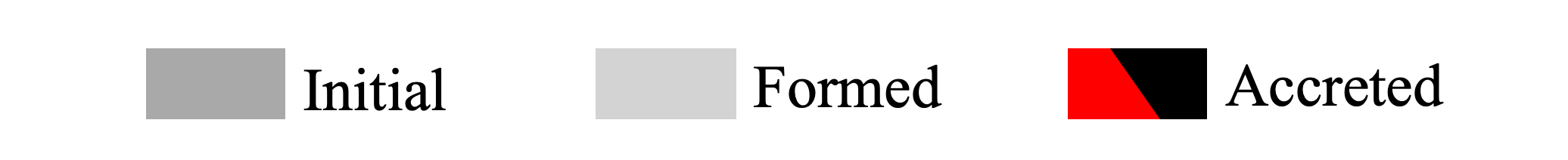}
    \caption{Contributions of accreted and formed stars to the composition of sixteen example clusters, at the end of our simulations. The examples are the four most massive clusters from the B-P2, B-P3, S-R2 and S-R3 simulations.  Dark grey wedges denote the initial (retained) stellar mass and light grey wedges denote the stellar mass formed in the cluster and retained to the last snapshot. Red (in clusters with primordial binaries) and black (in clusters without primordial binaries) wedges denote the accreted stellar mass that is retained to the last snapshot. }
    \label{fig:pie_charts}
\end{figure}

In Figure~\ref{fig:pie_charts}, we compare the relative impact of accretion and star formation on the assembly of our simulated embedded clusters. As examples, we show the four most massive clusters from the B-P2, B-P3, S-R2 and S-R3 simulations, which were run respectively to 2.0, 2.1, 2.0, and 2.2 Myr after the onset of star formation.  In all cases except the second most massive cluster in S-R2, the formation of new stars within the cluster contributes more mass to the cluster than the accretion of already-formed stars. In that example, accreted mass contributes 53\% of the total final mass of the cluster. The most extreme example of splitting is shown in the third most massive cluster in S-R3: the cluster split from the most massive cluster in the simulation 0.1 Myr before the last snapshot, and had a mass of 327~M$_{\odot}$ just after splitting (see initial mass of the third cluster in the fourth row of Figure~\ref{fig:pie_charts}). 

\begin{figure}
    \centering
    \includegraphics[width=\linewidth, clip=True, trim=0cm 0cm 0cm 0cm]{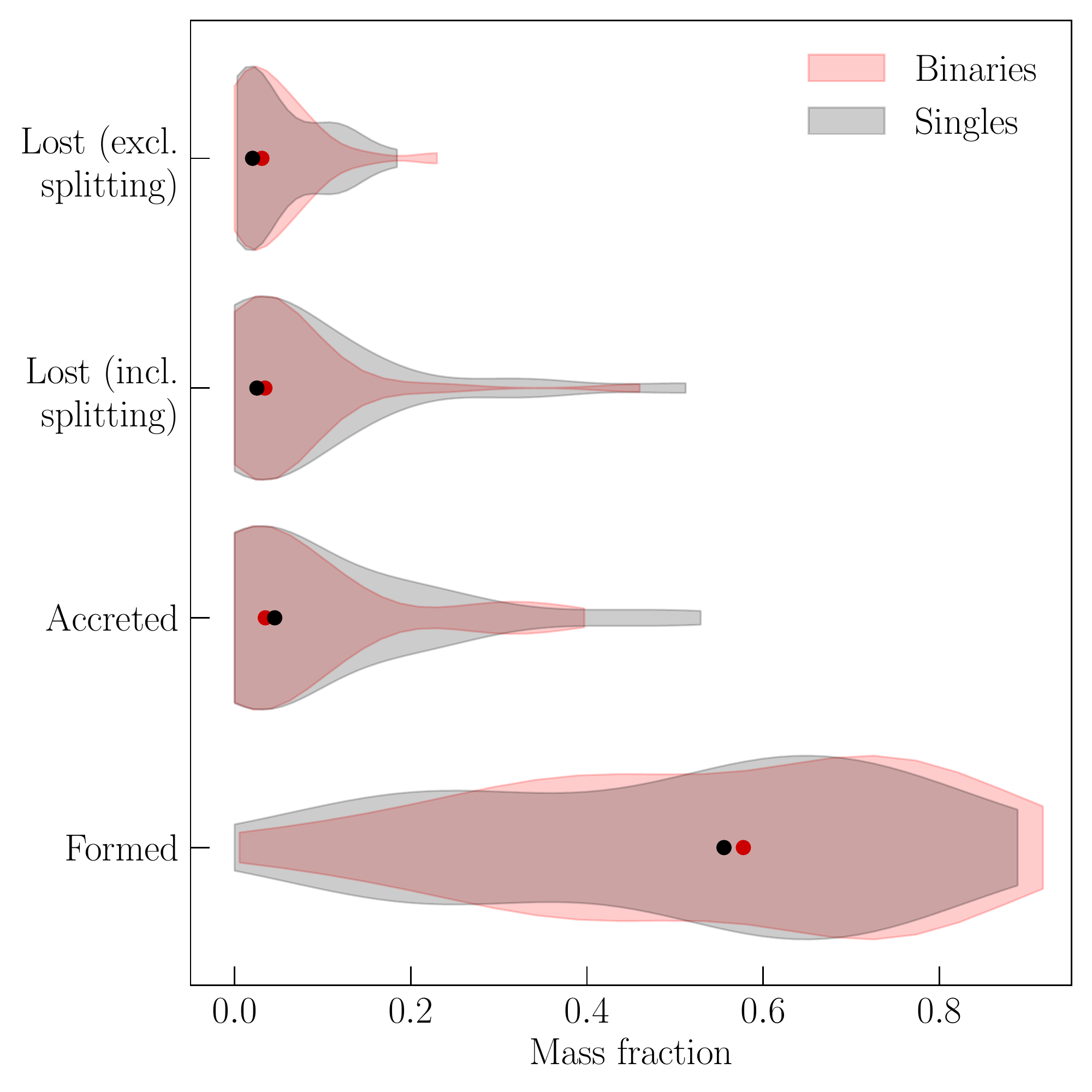}
    \caption{Distributions of mass fractions of lost, accreted, and formed stars for simulations with primordial binaries (\textit{Binaries}) and without primordial binaries (\textit{Singles}). Medians are shown as solid dots. Note that the accreted and formed fractions do not add up to 100\%, as neither includes the stellar mass present in the cluster when it is first identified.}
    \label{fig:violin}
\end{figure}

We present in Figure~\ref{fig:violin} an overview of the relative contributions of mass loss, accretion, and in-cluster star formation to the history of the embedded clusters in our simulations. The lost mass is calculated from the ratio of the mass lost by the cluster to the total mass acquired by the cluster over its history -- i.e. the final mass plus the lost mass. The accreted (formed) fraction is calculated as the fraction of the final stellar mass of the cluster that was accreted (formed) after the cluster was first identified. The accreted and formed fractions for a cluster therefore do not add up to 100\%, as the stellar mass present in the cluster when it is first identified also contributes. The first violin plot for the lost fractions does not include clusters that split into two clusters surviving for more than 0.1 Myr. 

There are no statistically significant differences in the final compositions of clusters with and without primordial binaries, as verified by a series of two-sample KS tests comparing the fractions of the stellar mass lost, accreted, and formed within the cluster for simulations with and without primordial binaries. We also find no statistically significant difference for clusters formed in simulations with $\alpha=0.8$, $\alpha=2.0$, and $\alpha=4.0$. We however find a rich variety of relative contributions from accretion and star formation, at all cluster masses. In other words, we find that there is no single dominant growth mechanism for clusters while they are still deeply embedded and actively star-forming, although generally stars formed \textit{in situ} outnumber accreted stars in a given cluster. 

\begin{figure}
    \centering
    \includegraphics[width=\linewidth, clip=True, trim=0cm 0.5cm 1cm 2cm]{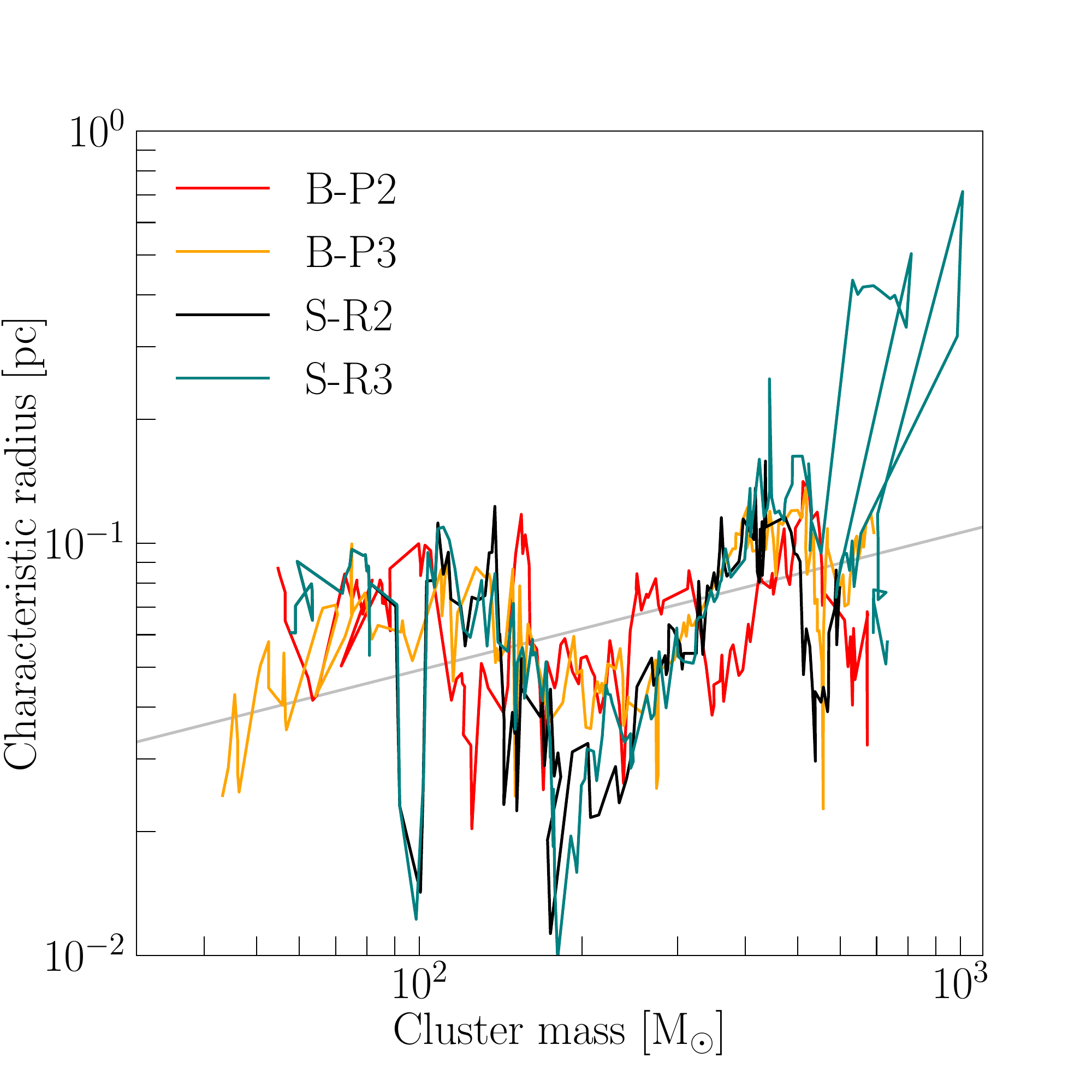}
    \caption{Characteristic radius $r_{50}$ of the most massive cluster in B-P2, B-P3, S-R2, and S-R3, against its mass. Each line represents the time evolution of a single cluster, with the leftmost end corresponding to the characteristic radius and mass when the cluster is first identified. The grey diagonal line corresponds to a constant density of 10$^5$ M$_{\odot}$pc$^{-3}$.}
    \label{fig:radius_mass}
\end{figure}

We have shown in Section~\ref{subsec:results_clusters} that cluster radius and cluster mass are uncorrelated for our full population of simulated embedded clusters. We now investigate how the radii of individual clusters change as they grow in mass. In Figure~\ref{fig:radius_mass}, we present as examples the evolution in radius-mass space of the most massive clusters in the B-P2, B-P3, S-R2 and S-R3 simulations. We find once again no correlation between radius and mass. We however note that radius can change by up to one order of magnitude without significant changes to the mass (see e.g. S-R3).  

\begin{figure*}
    \centering
    \includegraphics[width=\linewidth, clip=True, trim=5cm 5cm 8cm 5cm]{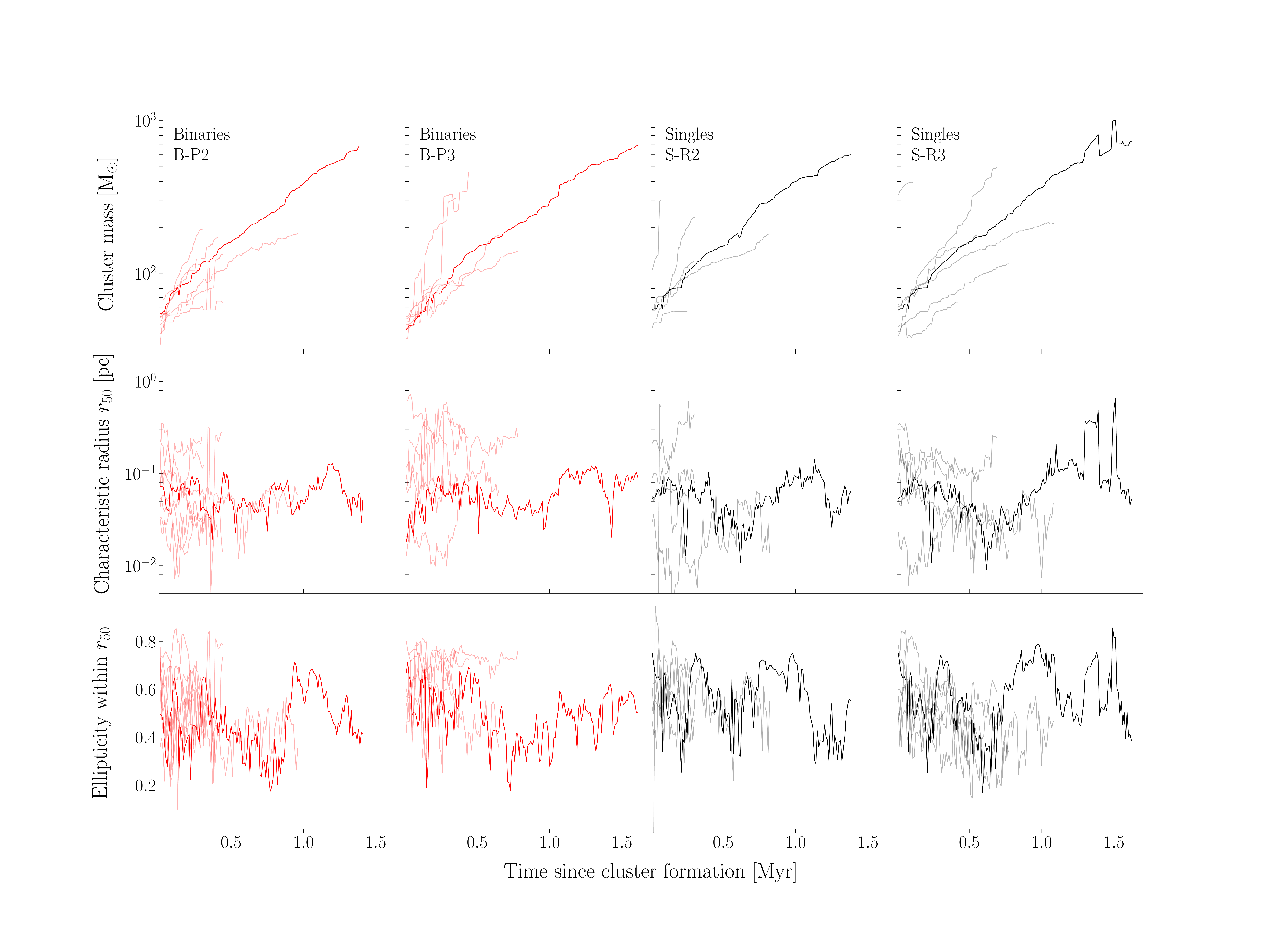}
    \caption{Mass (top), characteristic radius $r_{50}$ (middle), and ellipticity within the characteristic radius (bottom) of individual clusters in B-P2, B-P3, S-R2, and S-R3 against the time since their formation. The most massive cluster in each simulation is shown in bold; simulations with primordial binaries are shown in red and simulations without primordial binaries are shown in black.}
    \label{fig:properties_time}
\end{figure*}

We also investigate the evolution of the mass, radius, and ellipticity of individual clusters as a function of the time since they were first identified. In Figure~\ref{fig:properties_time}, we plot these quantities against the time since cluster formation for all clusters identified in B-P2, B-P3, S-R2, and S-R3. The first important result we glean from this plot is that the mass growth of our simulated clusters is not always monotonic. A clear example is the most massive cluster in S-R3, which loses $\gtrsim 300$~M$_{\odot}$ within 0.01 Myr as it splits, as discussed above. In most cases, however, the mass of the most massive cluster tends to grow exponentially with time. In all cases except for S-R1, the most massive cluster is also the longest-lived cluster. In most cases, it is however not the one with the highest growth rate, which suggests that another cluster could become more massive at later times. Overall, simulations with and without primordial binaries follow the same general trends. 

In Figure~\ref{fig:properties_time}, we also explore the time evolution of the characteristic radius $r_{50}$ of the clusters in our simulations. We find no correlation between the characteristic radius of a cluster and the time since it was formed. This is an important result as it suggests that the evolution of our embedded simulated clusters is not yet dominated by their internal dynamics, which should cause expansion~\citep[see e.g.][for recent simulations]{Torniamenti2021}. We further highlight that considerable changes in cluster radius, of half an order of magnitude, occur on timescales shorter than 0.01~Myr (i.e. between two consecutive snapshots). We also plot the ellipticity of the distribution of cluster stars enclosed within their characteristic radius $r_{50}$ against time since cluster formation. We once again find no correlation, and find that considerable changes can take place over $\sim 0.01$ Myr. Changes in cluster size and shape, while the clusters are actively forming, are therefore driven by physical processes more complex than simply growth in mass or effects from stellar dynamics. 

\begin{figure*}
    \centering
    \includegraphics[width=0.7\linewidth, clip=True, trim = 1cm 3.5cm 1.5cm 4.5cm]{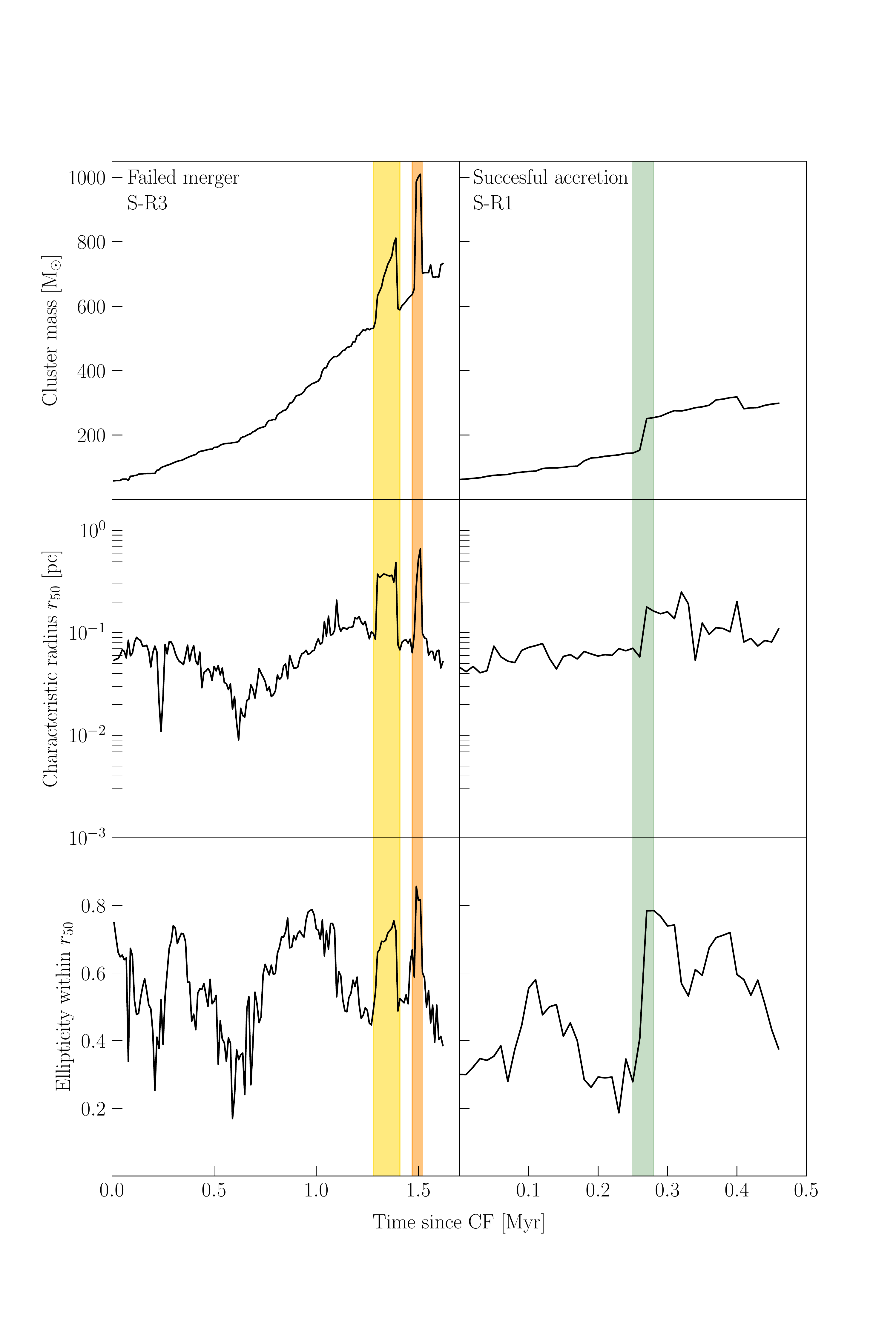}
    \caption{Morphology histories for the most massive clusters in S-R3 (left) and S-R1 (right).The two regions highlighted in the left panel demonstrate a failed merger. The sudden changes in the stellar mass coincide with very sharp changes in radius and ellipticity. The region highlighted in the right panel denotes a successful accretion event. The growth in mass corresponds to a growth in characteristic radius and in ellipticity. In contrast with the left panel, the radius and ellipticity do not decrease sharply immediately after the event: since the accretion was successful, they decrease more smoothly over the next 0.2 Myr.}
    \label{fig:cluster_histories}
\end{figure*}

Rapid changes in morphology are driven by accretion and splitting events. We compare the timing of changes in cluster mass, characteristic radius, and ellipticity in Figure~\ref{fig:cluster_histories}, and find that they occur at the same times. In particular, we find that the times for the local minima and maxima in characteristic radius and ellipticity match. This is not due to a general correlation between size and shape, as shown in Figure~\ref{fig:radius_ellipticity}: rather, it indicates that clusters grow more elliptical and grow in size at the same time, when they are actively accreting an infalling group of stars. "Failed" accretion events, or events followed by a splitting of the cluster, result in a rapid increase of the radius and ellipticity followed by a rapid decrease of the radius and ellipticity as the cluster returns to its original state (see left panel of Figure~\ref{fig:cluster_histories}). When the accretion event is successful, the cluster's radius and ellipticity also grow rapidly but decrease more smoothly after the event (see right panel of Figure~\ref{fig:cluster_histories}). 

Together, our results indicate that the early structure evolution of embedded clusters is driven by processes arising from the larger cluster-forming region -- such as a burst of star formation due to an inflow of gas, or the accretion of a group of stars -- rather than by their internal dynamics. In particular, we find that the composition of the clusters is being modified by the formation of new stars or the accretion of already formed stars over timescales much shorter than the clusters' relaxation time. We can obtain back-of-the-envelope estimates of the lower and upper limit on the relaxation times for the clusters in our simulations with 
\begin{equation}\label{eq:t_relax}
    t_{\text{relax}} \simeq \frac{0.1N}{\ln N} t_{\text{cross}}
\end{equation}
where $t_{\text{cross}}$ is the crossing time,
\begin{equation}\label{eq:t_cross}
    t_{\text{cross}} \simeq \frac{1}{\sqrt{G\rho}}
\end{equation}
and $\rho$ is the average density inside a particle's orbit~\citep{GalacticDynamics}. Although both equations are only exact for spherical systems, they nonetheless provide us with a simple estimate of the timescales relevant for our simulated embedded clusters. For the lower limit on the relaxation time, we assume a cluster with 100 members, the minimum possible in our analysis framework, and use a density of 10$^{4}$~M$_{\odot}$~pc$^{-3}$, which is towards the low end of our density values but still common. For the upper limit, we assume 1000 members, which is at the high end for our simulated embedded clusters, and use a density of 10$^{5}~$M$_{\odot}$~pc$^{-3}$. These give us estimated relaxation times of $\sim 0.32$ Myr and $\sim 0.68$ Myr, which are longer than the timescales over which the mass -- and therefore the number of stars -- of the embedded clusters change. Indeed, sudden accretion events, like the attempted merger shown in the left panel of Figure~\ref{fig:cluster_histories}, may change a cluster's characteristic radius $r_{50}$ by up to an order of magnitude, and its mass by a factor of $\sim~1.5$, over $\sim~0.01$ Myr. Those timescales are more than 10 times shorter than the estimated relaxation times for the clusters. We conclude that the embedded clusters present in our simulations are not relaxed, despite their small sizes, due to how frequently they form or accrete new stars. Their dynamical evolution is therefore still driven by the overall gravitational potential of the simulation domain, dominated by the gas, rather than by two- or few-body encounters within the cluster. 

\section{Discussion}\label{sec:Discussion}
We now discuss the results presented in Section~\ref{sec:Results}. We compare the efficiency of star formation in our simulations to recent observations and simulations of cluster-forming regions. We also discuss the broader implications of our results for observational and computational studies of embedded clusters. We end by outlining areas for future work. 

\subsection{Star Formation Efficiency} 
We have explored a range of realistic initial virial parameters $\alpha$, ranging from a low virial parameter typical of more massive clouds in which young massive clusters (YMCs) form to a moderately high virial parameter typical of the 10$^4$ M$_{\odot}$ clouds in the solar neighbourhood. The general trend is for our simulations with higher $\alpha$ to have lower SFE and a smaller mass for the most massive cluster, in agreement with observations~\citep[e.g.][]{Schruba2019} and GMC-scale hydrodynamics simulations~\citep[e.g.][]{Howard2016}. The physical quantities obtained for the full simulated domain and for individual and average clusters are generally in agreement with observations of Galactic star-forming regions and young clusters. The integrated SFE after one free-fall time is about 1\% in our simulations with $\alpha=4.0$ while it is about 3\% in our simulations with $\alpha=2.0$. Observations suggest a SFE per free-fall time of about 1\% on pc-scale clouds, with scatter up to about 3\%~\citep[][and references therein]{Krumholz2019}; this is consistent with the results from our simulations. 

The integrated SFEs after one free-fall time in our simulations are also consistent with the results from the \textsc{STARFORGE} simulations conducted by~\citet{Guszejnov2022} with a cloud mass of 2~x~10$^4$~M$_{\odot}$ and virial parameters $\alpha=1.0, 2.0, 4.0$. Those simulations include models for protostellar outflows, radiation, stellar winds, and supernovae. Our simulations with $\alpha=0.8$ have SFEs per free-fall time $\lesssim$ 10\%, which is similar to what they obtain in their simulations with $\alpha=1.0$. It is further possible to compare our simulations with $\alpha=0.8$ to the work conducted by~\citet{Howard2016} using \textsc{FLASH} with radiative feedback. The SFR after one free-fall time is $\sim$~10$^{-3}$~M$_{\odot}$ yr$^{-1}$ for our 10$^4$~M$_{\odot}$ clouds with $\alpha=0.8$. This is consistent -- after scaling for cloud mass -- with the SFRs obtained by~\citet{Howard2016} for their clouds with $\alpha=0.5$ and $\alpha=1.0$, where the SFR after one free-fall time is $\sim$~10$^{-1}$~M$_{\odot}$~yr$^{-1}$ for a 10$^6$~M$_{\odot}$ cloud. 

\subsection{Implications for Observations}
There are two key takeaways from our simulations that can be applied to observed embedded clusters. First, the structure of embedded clusters -- such as shape and size -- can change considerably over timescales as short as 0.01~Myr, due to new star formation or accretion. Those changes are not monotonic, do not follow a general trend, and are not driven by internal dynamics. This contrasts with studies of the early evolution of gas-free young star clusters. \citet{Torniamenti2021}, for example, find that gas-free young clusters expand faster in the presence of primordial binaries. The evolution of the morphology of our embedded clusters, however, is driven primarily by the acquisition of news stars via star formation or accretion. Both processes are themselves driven by gas dynamics: star formation takes place within dense, converging flows of gas, while the accretion of already-formed stars is driven by the gravitational dynamics of the gas, which still accounts for $\gtrsim~70\%$ of the mass within the simulation domain at the end of the runs. The simulations presented here focus on the deeply embedded stages of cluster formation. Tentative conclusions about the behaviour of our clusters up to and following gas expulsion can be reached by considering the results from the simulations that we presented in~\citet{Lewis2022}, albeit with some caveats: the simulations presented in~\citet{Lewis2022} have a spatial resolution four times coarser than the present work (0.27 pc versus 0.0683 pc), consider only one virial parameter $\alpha$ selected to promote abundant star formation, and do not include binaries.

We stress therefore that the observed state of Galactic embedded clusters in their first stages of formation is instantaneous. We argue that no conclusions about the future evolution of a very young embedded cluster can be drawn from its current size or ellipticity: the cluster's current state gives no information about whether the radius or ellipticity will increase or decrease in the future. Environment and recent changes in stellar mass play a role at least as important as internal processes such as two- or few-body encounters in setting embedded clusters' dynamical states. We further note that more information about the stellar content and kinematics of very young embedded clusters --including information about binaries -- would not allow us to predict their evolution better. We thus also predict that observations of stellar positions and velocities -- that could be used to verify boundedness, investigate cluster expansion, and measure cluster shape -- cannot be used to infer the presence of a significant number of binaries in embedded clusters. 

Second, we find that clusters tend to have large ellipticities and large characteristic radii when they are accreting new stars. Examples are shown in Figure~\ref{fig:cluster_histories}. A large ellipticity for an embedded cluster with a large radius, that persists despite projection effects, could be clear observational evidence that the embedded cluster is currently accreting -- or has recently accreted -- new stars without requiring any stellar velocity data.~\citet{Torniamenti2021}, in their simulations of the early evolution of gas-free young clusters, similarly find that clusters in the process of merging appear more elongated.
We thus argue that the size and shape of observed embedded clusters can inform our understanding of their recent history but not of their future evolution.
 
\subsection{Implications for Larger-Scale Simulations} 
Simulations of YMC formation with hydrodynamics and stellar feedback require very high gas masses for the initial GMC (three orders of magnitude above what we consider here, around 10$^7$ M$_{\odot}$) and thus often model sub-grid clusters with sink particles that can grow in mass by merging with other sinks and accreting gas~\citep[e.g.][]{Howard2016, Howard2018}. \citet{Karam2022} have highlighted some of the limitations of this model, by showing that collisions between clusters do not always result in a single, merged cluster and that even when they do, the bound mass of the resulting cluster is less than the sum of the bound masses of the progenitors. They also find that cluster radii grow following a merger. We reinforce here those conclusions, and further note that groups of recently formed stars identified as cluster members -- that would form within a sub-grid cluster sink -- can escape a cluster and can even be identified as a new cluster later in the simulation if they escape together. In particular, clusters can lose up to $\sim 50\%$ of their stellar mass if they split, and up to $\sim 30\%$ without splitting. A significant fraction of the stellar mass formed or accreted by a cluster can be lost on pre-supernova timescales, which is not accounted for in cluster sink models. 

Our embedded clusters tend to build up their mass mostly by forming new stars within the cluster, although they can accrete up to $\sim 50\%$ of their mass in already formed stars. Both processes contribute to the clusters' growth in mass on timescales much shorter that the clusters' relaxation times. The dynamical evolution of the clusters remains driven by gravitational processes on the scale of the full simulation domain, such as the collapse of the gas, rather than by internal processes. This is a plausible cause of the diversity of cluster histories within the same simulation, as each individual cluster forms in a different local environment. \citet{Howard2016} found a similar spread for their cluster sinks: for their clusters in the 10$^2$-10$^3$ M$_{\odot}$ mass range, similar to our simulated embedded clusters, they find that between 0\% and $\sim$ 60\% of the clusters' stellar mass is accreted.

Approximating embedded clusters as relaxed, spherical collections of gas and stars does not give an accurate representation of the clusters' dynamical state. Furthermore, using spheres as a proxy for the shape of embedded clusters -- or as a tool to measure cluster size, e.g. from Lagragian radii -- may not be appropriate, as our clusters generally have ellipticities around 0.5, which indicates a factor of 2 difference between the major and minor axes of the stellar distribution.

\subsection{Directions for Future Work}\label{subsec:Future}
The simulation time for which we can evolve our models is currently limited by the high computational cost associated with following the dynamics of a large number of close binaries concurrently with radiative transfer and hydrodynamics. Including a self-consistent treatment of binary dynamics in simulations of embedded clusters as they reach gas expulsion is however essential to advancing our understanding of how star clusters form in galaxies. Although our results here indicate that gas dynamics dominate in the deeply embedded phase of cluster formation, the effects of binaries on the dynamics of clusters during gas expulsion remain unknown, and binaries are known to have an important impact on the evolution of gas-free clusters~\citep[e.g.][]{Heggie1975, Hills1975, Torniamenti2021}. Pursuing similar simulations with a large number of close binaries over timescales sufficient to reach gas expulsion is therefore our next goal. This will require the use of a different N-body and few-body solver, to replace \textsc{Ph4} and \textsc{Multiples}.

Directions for future work also include improvements to the treatment of gas and the stellar feedback in our simulations. Magnetic fields are not used in the current work due to their high computational cost. Future work will include comparisons of simulations with and without magnetic fields, as they are known to participate in the regulation of star formation~\citep{Price2008}. We also note that the amount of mass injected by wind feedback in our simulations is an upper limit, since the~\citet{Vink2000} prescription for mass loss rates is likely too high by a factor of $\sim$3~\citep{Smith2014} and our winds are mass-loaded to avoid extremely short timesteps~\citep{Wall2020}. The shock fronts in compact colliding wind binaries are not resolved due to our gas spatial resolution, such that we underestimate the heating from the winds. Any modulation of the feedback coming from interacting binaries is neglected. Including feedback from binaries is non-trivial, but is something we hope to address in future work. Our simulations also currently do not include protostellar jets and outflows. We expect the caveats outlined above to affect the spatial distribution of the feedback in our simulations, but not to significantly under- or overestimate the overall feedback budget.

\section{Summary}\label{sec:Conclusion}
We have conducted a suite of hydrodynamics simulations of star cluster formation with a state-of-the-art treatment of stellar dynamics down to the scale of individual binaries, as well as active star formation via sink particles, and stellar feedback. We have explored a range of realistic initial virial parameters $\alpha=0.8, 2.0, 4.0$, at a fixed initial cloud mass of 10$^4$ M$_{\odot}$, five different models for the formation (or not) of primordial binaries, and seven different random seeds for stochastic star formation. Most of our simulations have progressed to 2.0 Myr after the onset of star formation, which is the same as the timescales we considered in~\citet{Cournoyer-Cloutier2021}. This allows us to investigate the relative impacts of the cloud-scale gas environment and internal two- or few-body dynamics while gas dynamics are still dominated by the gravitational collapse of the gas and not yet by the effects of stellar feedback. We have used a combination of tools to identify and characterize clusters, and arranged our analysis around three main axes: the properties of the full simulation domain (Section~\ref{subsec:results_overview}), the properties of the identified clusters as a population (Section~\ref{subsec:results_clusters}), and the time evolution of individual clusters (Section~\ref{subsec:results_time}). We have verified that the SFE of our simulation domains, as well as the sizes, densities, and ellipticities of our embedded clusters are consistent with observations. 

We explored the relative impact of the cloud's initial virial parameter $\alpha$ and stellar dynamics (using the presence of primordial binaries as a proxy) on cluster structure and evolution. We have found the following:
\begin{enumerate}
    \item The choice of initial virial parameter $\alpha$ has the largest systematic effect on the global properties of the simulation domain, such as the SFR and SFE.
    \item The presence of primordial binaries or individual massive stars causes scatter in the SFR and SFE, but no systematic effect. The scatter is smaller than the systematic effects caused by changes $\alpha$. Stochastic effects from individual stars are important due to the low cluster masses ($\lesssim 1000$ M$_{\odot}$) considered in our simulations. 
\end{enumerate}
Our simulated embedded clusters are not relaxed, as their mass changes due to accretion or star formation on timescales significantly shorter than their relaxation times. We thus find that their dynamical evolution is driven by the local gravitational potential (from the gas and stars) rather than by two- or few-body encounters (and therefore the presence of binaries). We have also tracked how cluster structure evolves during the earliest stages of cluster formation. We find considerable variation in cluster histories; examples are shown in Figure~\ref{fig:pie_charts}. We summarize our results on cluster evolution as follows:
\begin{enumerate}
    \setcounter{enumi}{2}
    \item Cluster mass generally grows through star formation rather than accretion, although some individual clusters acquire up to half of their final mass by accretion.
    \item The mass of individual clusters generally grows exponentially, although this growth is not monotonic. Clusters can lose up to half of their mass while they assemble.
    \item The size, density, and ellipticity of clusters does not follow any particular trend as the cluster acquires more mass. Changes in size, density, and ellipticity can take place over timescales as short as 0.01~Myr.
    \item Recent accretion coincides with simultaneous sharp increases in characteristic radius and ellipticity. We propose that observed embedded clusters with high ellipticities are in the process of accreting stars.
        
\end{enumerate}
The earliest stages of star cluster formation, when stars are still embedded in their natal gas and stars are still actively forming, are driven by a variety of competing physical processes; the structure of embedded star clusters changes quickly. We caution observers that the state in which an embedded cluster is observed is instantaneous. Over the timescales considered in this work, cluster dynamical evolution is driven by the overall gravitational potential of the star-forming region, as individual clusters acquire new stars on timescales much shorter than their relaxation times. 

\section*{Acknowledgements}
We warmly thank Marta Reina-Campos and James Wadsley for useful discussion. CCC is supported by a Canada Graduate Scholarship -- Doctoral from the Natural Sciences and Engineering Research Council of Canada (NSERC). CCC also acknowledges funding from a Queen Elizabeth II Graduate Scholarship in Science and Techonology (QEII-GSST) for the 2021-2022 academic year. AS and WEH are supported by NSERC. SA is grateful for funding from NSF grant AST-2009679. BP is supported through a fellowship from the International Max Planck Research School for Astronomy and Cosmic Physics at the University of Heidelberg (IMPRS-HD). AT was partly supported by NASA FINESST 80NSSC21K1383. AT was also supported through a NASA Cooperative Agreement awarded to the New York Space Grant Consortium.  AT, MMML and SLWM were partly supported by NSF grant AST18-15461. SLWM was also supported by NSF grant AST18-14772. MW acknowledges funding from NOVA under project number 10.2.5.12.

The simulations were conducted on the Digital Research Alliance of Canada supercomputer Graham under the resource allocation RRG \#4398: \textit{The Formation of Star Clusters in a Galactic Context}, and on the supercomputer Cartesius under the Dutch National Supercomputing Center SURF grant 15520. In addition to the software cited in-text, we have made use of the matplotlib~\citep{matplotlib}, numpy~\citep{numpy}, scipy~\citep{scipy} and yt~\citep{yt} Python packages for plotting and analysis.   

\section*{Data Availability}
The data underlying this article will be shared on reasonable request to the corresponding author.



\bibliographystyle{mnras}
\bibliography{bibliography} 




\appendix

\section{Binary Prescriptions}\label{secA:binaries}
\begin{enumerate}
    \item \textit{Field distribution} \,\,\, This is our fiducial distribution, based on statistics for all companions to main sequence stars in the Galactic field. It is presented in detail in~\citet{Cournoyer-Cloutier2021} and is based on observations compiled by~\citet{Moe2017} and~\citet{Winters2019}.
    \item \textit{10\% random pairing} \,\,\, This prescription is based on that used in~\citet{Sills1999}; similar prescriptions continue to be used in current state-of-the-art N-body or Monte-Carlo simulations of massive star clusters~\citep[see e.g.][]{Kamlah2022, Wang2022}. It imposes a mass-independent binary fraction of 10\%, with a period drawn from a flat distribution in $\log P$ (between 0.5 and 7.5, in days), and an eccentricity drawn from a thermal distribution. This model tends to under-produce binaries compared to the Galactic field, but nonetheless contains low-mass binaries that do not form naturally in models without primordial binaries. 
    \item \textit{100\% random pairing} \,\,\, This prescription is the same as the one described above, with a binary fraction of 100\% at all masses. 
    \item \textit{Field distribution for $M < 0.6 M_{\odot}$ and no close massive binaries} \,\,\, This prescription is also based on the algorithm presented in~\citet{Cournoyer-Cloutier2021}, but shifts all the periods to higher values $\tilde{P}$ for stars with masses above $0.6$ M$_{\odot}$ following $\tilde{P} = 10^P$, where $P$ is the period drawn from the algorithm. The specific choice of period shift is motivated by a typo we found in the binary generation algorithm we used in~\citet{Cournoyer-Cloutier2021}, which caused us to draw $\log P$ instead of $P$ from our observations-based distribution. This typo does not affect the conclusions of the previous paper, as those were drawn from comparisons to the distribution of formed binaries (and not from comparisons to observations).
\end{enumerate}

\section{Ellipsoids from Inertia Tensors}\label{secA:inertia}

\begin{figure*}
    \centering
    \begin{minipage}{.48\linewidth}
        \centering
        \includegraphics[width=\linewidth, clip=True, trim=1cm 1cm 1cm 1cm]{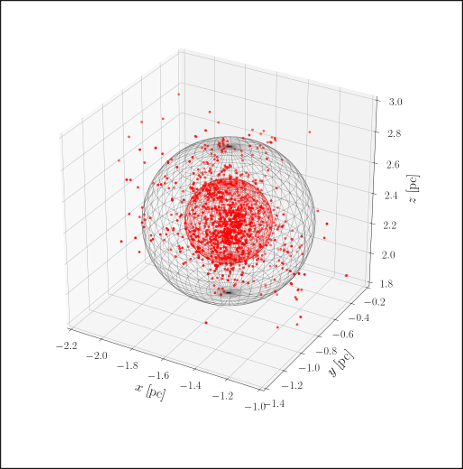}
    \end{minipage}%
    \begin{minipage}{0.48\linewidth}
        \centering
        \includegraphics[width=\linewidth, clip=True, trim=1cm 1cm 1cm 1cm]{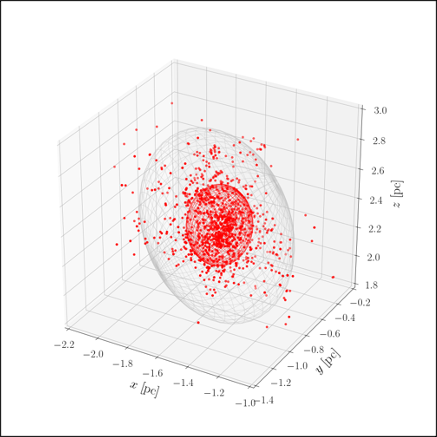}
    \end{minipage}
    \caption{3D spheres (left) and ellipsoids (right) enclosing 50\% (red) and 90\% (grey) of the stellar mass of the example cluster, taken from S-R3. All individual stars bound to the cluster are shown in red.}
    \label{fig:sphere_ellipsoid_example}
\end{figure*}

We present in Figure~\ref{fig:sphere_ellipsoid_example} an example of 3D ellipsoidal surfaces enclosing 50\% and 90\% of the cluster mass, compared to the 50\% and 90\% Lagragian radii for the same stellar distribution. We use the reduced inertia tensor~\citep{Thob2019},
\begin{equation}\label{eq:inertia}
    \textbf{I} = \begin{pmatrix}
            \text{I}_{xx} & \text{I}_{xy} & \text{I}_{xz}\\
            \text{I}_{xy} & \text{I}_{yy} & \text{I}_{yz}\\
            \text{I}_{xz} & \text{I}_{yz} & \text{I}_{zz}
        \end{pmatrix}
\end{equation}
where the individual elements $I_{ij}$ are calculated from 
\begin{equation}\label{eq:inertia_element}
    \textbf{I}_{ij} =  \frac{\sum_{a} \Bigg( \frac{ m_a\Big(\mathbb{I} r_a^2\Big)_{ij} - \Big(\vec{x}_{a}\Big)_{i}\Big(\vec{x}_{a}\Big)_{j}}{r_a^2}\Bigg)}{\sum_{a} \Big(\frac{m_a}{r_a^2}\Big)} 
\end{equation}
and 
\begin{equation}
    r_a^2 = \vec{x}_a \cdot \vec{x}_a
\end{equation}
where $\vec{x}_{a}$ is the vector distance from star $a$ to the cluster's centre of mass.
The reduced inertia tensor minimizes the impact of stars in the outskirts of the cluster on the calculated shape. We obtain the principal axes $a$, $b$ and $c$ from the eigenvalues $\lambda_i$ of the reduced inertia tensor, such that  
\begin{equation}\label{eq:eigenvalues}
\begin{aligned}
    \lambda_a &\propto b^2 + c^2 \\
    \lambda_b &\propto a^2 + c^2 \\
    \lambda_c &\propto a^2 + b^2. 
\end{aligned}
\end{equation}
Solving this system of equations, we recover initial guesses for the principal axes
\begin{equation}\label{eq:axes}
\begin{aligned}
    a &\propto \sqrt{-\lambda_a + \lambda_b + \lambda_c} \\
    b &\propto \sqrt{\lambda_a - \lambda_b + \lambda_c} \\
    c &\propto \sqrt{\lambda_a + \lambda_b - \lambda_c}. 
\end{aligned}
\end{equation}
The initial guesses from Equations~\ref{eq:eigenvalues} and~\ref{eq:axes} are then rescaled iteratively to enclose 50\% or 90\% of the stellar mass. 
We adopt the idea of iterative fitting from~\citet{Thob2019}, and adapt the 2D code from~\citet{Hill2021} to handle 3D distributions. The main steps of the fitting algorithm are as follows: 
\begin{enumerate}
    \item Identify the centre of mass of the cluster; 
    \item Take the stars enclosed within a given Lagrangian radius and get the shape for this distribution from the reduced inertia tensor in Equations~\ref{eq:inertia} and~\ref{eq:inertia_element};
    \item Increase or decrease the size of the ellipsoid until the required (50\% or 90\%) fraction of the mass is enclosed;
    \item Recalculate the shape from the inertia tensor associated with the stars now enclosed in the ellipsoid;
    \item Repeat the steps above until the change in shape is less than a given tolerance between two iterations.
\end{enumerate}
The default tolerance is 1\% but we raise it to 2\% for systems with fewer than 500 (but at least 200) stars and to 5\% for systems with fewer than 200 stars. We do not fit an ellipsoid when the most massive star in a cluster accounts for more than 50\% of its mass. We encounter this situation for one cluster in a few consecutive snapshots in S-R2.



\bsp	
\label{lastpage}
\end{document}